\documentclass[12pt]{article}

\newenvironment{wileykeywords}{\textsf{Keywords:}\hspace{\stretch{1}}}{\hspace{\stretch{1}}\rule{1ex}{1ex}}

\usepackage{amsmath,amssymb}
\usepackage{graphicx}
\usepackage{color}
\usepackage{dcolumn}
\usepackage{bm}
\usepackage[numbers,super,comma,sort&compress]{natbib}
\usepackage{multirow}
\usepackage{booktabs}
\usepackage{makecell}
\usepackage{threeparttable}
\usepackage{subfigure}
\usepackage{tikz}
\usepackage{overpic}
\usetikzlibrary{shapes.geometric, arrows, positioning}
\usepackage{caption}

\usepackage{algorithm}
\usepackage{algorithmic}
\usepackage{xargs} 
\usepackage{lipsum}  
\usepackage[colorinlistoftodos,prependcaption,textsize=tiny]{todonotes}

\newcommandx{\addcite}[2][1=]{\todo[linecolor=red,backgroundcolor=red!25,bordercolor=red,#1]{#2}}
\newcommandx{\checkthis}[2][1=]{\todo[linecolor=lime,backgroundcolor=lime!25,bordercolor=lime,#1]{#2}}
\newcommandx{\addsome}[2][1=]{\todo[linecolor=green,backgroundcolor=green!25,bordercolor=green,#1]{#2}}
\newcommandx{\modifythis}[2][1=]{\todo[linecolor=cyan,backgroundcolor=cyan!25,bordercolor=cyan,#1]{#2}}

\definecolor{background-color}{gray}{0.98}

\title{{Fault-tolerant} Quantum Chemical Calculations with Improved Machine-Learning Models}

\author{ Kai Yuan \footnotemark[3] \footnotemark[2] \footnotemark[5], 
Shuai Zhou \footnotemark[4] \footnotemark[6] \footnotemark[5],
Ning Li \footnotemark[8],
Tianyan Li \footnotemark[2],
Bowen Ding  \footnotemark[9], \footnotemark[1] \\
Danhuai Guo \footnotemark[3] \footnotemark[1],
Yingjin Ma  \footnotemark[2] \footnotemark[1]}

\footnotetext[5]{Authors contributed to this work equally}
\footnotetext[1]{Corresponding authors: \textit{dingbowen1214@163.com}, \textit{gdh@buct.edu.cn}, and \textit{yingjin.ma@sccas.cn} }
\footnotetext[3]{College of Information Science and Technology, Beijing University of Chemical Technology, Beijing 100029, China}
\footnotetext[2]{Computer Network Information Center, Chinese Academy of Sciences, Beijing 100190, China}
\footnotetext[6]{Institute of Computing Technology, Chinese Academy of Sciences, Beijing 100190, China}
\footnotetext[4]{School of Computer Science and Technology, University of Chinese Academy of Sciences, Beijing 101408, China}
\footnotetext[8]{College of Chemistry and Materials Engineering, Wenzhou University, Wenzhou 325035, China}
\footnotetext[9]{Institute of Chemistry, Chinese Academy of Sciences, Beijing 100190, China}

\begin{document}

\maketitle

\begin{abstract}
{Easy and effective usage of computational resources is crucial for scientific calculations. 
Following our recent work of machine-learning (ML) assisted scheduling optimization [Ref: \textit{J. Comput. Chem.} \textbf{2023}, 44, 1174], we further propose 1) the improved ML models for the better predictions of computational loads, and as such, more elaborate load-balancing calculations can be expected; 2) the idea of coded computation, i.e. the integration of gradient coding, in order to introduce fault tolerance during the distributed calculations; and 3) their applications together with re-normalized exciton model with time-dependent density functional theory (REM-TDDFT) for calculating the excited states.
Illustrated benchmark calculations include P38 protein, and solvent model with one or several excitable centers.   
The results show that the improved ML-assisted coded calculations can further improve the load-balancing and cluster utilization, owing primarily profit in fault tolerance that aims at the automated quantum chemical calculations for both ground and excited states. 
}
\end{abstract}

\begin{wileykeywords}
Coded computing, load-balancing, interacting energy, fragmented approach, exciton model.
\end{wileykeywords}

\clearpage


 \begin{figure}[h]
 \centering
 \colorbox{background-color}{
 \fbox{
 \begin{minipage}{1.0\textwidth}
 \includegraphics[scale=0.11,bb=0 0 3000 2500]{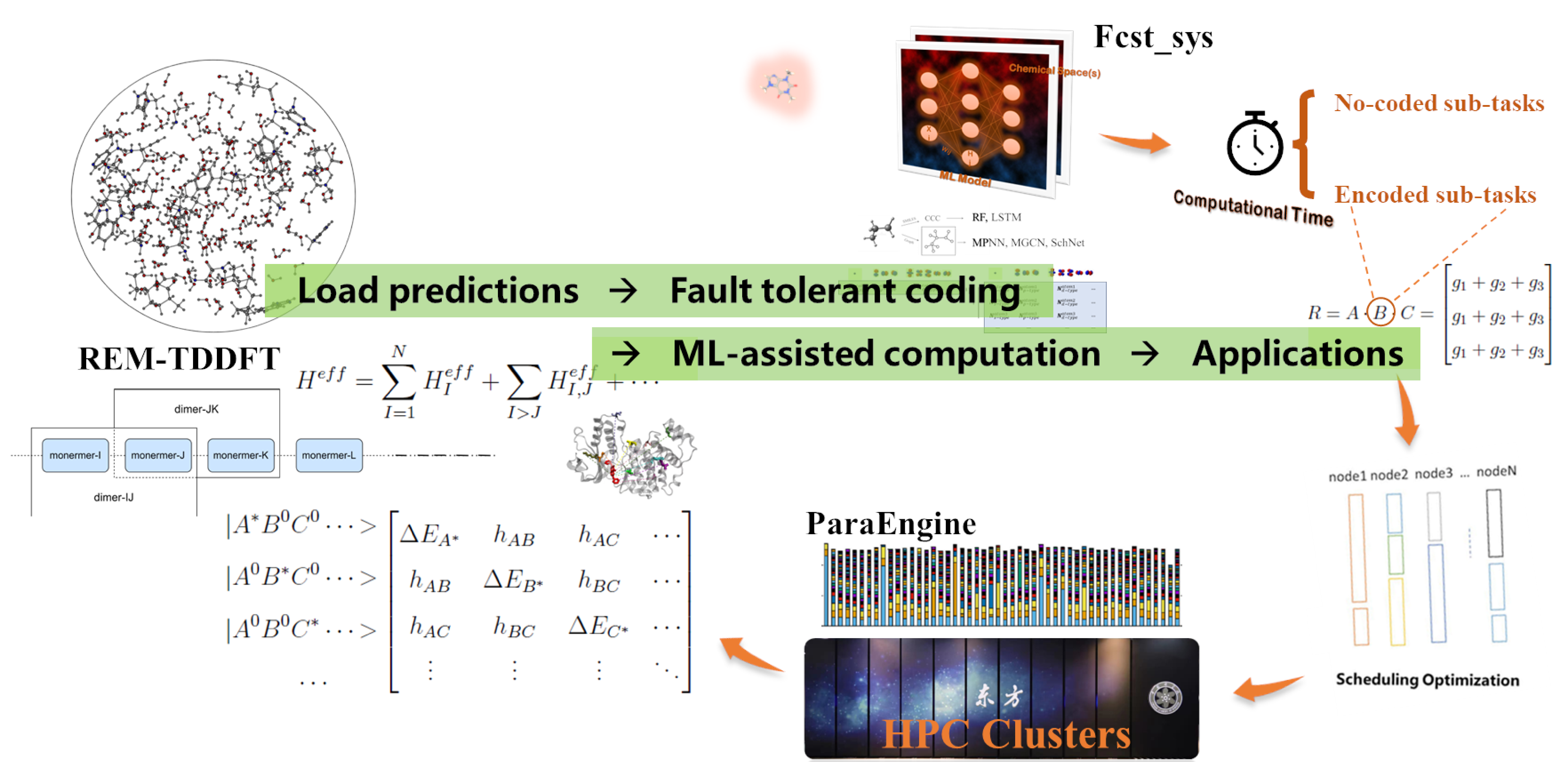} 

 (75 words.) 
{ We present a procedure for easy and effective implementations of coded quantum chemical calculations with improved machine-learning (ML) models. Employing this procedure, we show that the improved ML-assisted coded calculations can further improve the load-balancing and cluster utilization, owing primarily profit in fault tolerance that aims at the automated quantum chemical calculations for both ground and excited states.
 }
 \end{minipage}
 }}
 \end{figure}

  \makeatletter
  \renewcommand\@biblabel[1]{#1.}
  \makeatother

\renewcommand{\baselinestretch}{1.5}
\normalsize

\clearpage

\section{\sffamily \Large Introduction}

The {primary objective of quantum chemical calculations has always been to provide reliable descriptions of various properties for different molecular systems.} However, the complexity of traditional quantum mechanics (QM) methods, particularly the computation of electron repulsion integrals (ERIs) using atomic basis functions, limits their application {to large molecular systems}. Currently, there is a growing interest in applying quantum chemical calculations to the study of biological macromolecular systems. While QM methods are typically restricted to relatively small systems (tens of atoms), {their scope} can be expanded using fragment-based techniques
\cite{kitaura1999fragment,he2005new,he2006generalized,fedorov2009fragment,mayhall2011molecules,wang2013electrostatically,10.1002,doi:10.1021/acs.jpca.2c00601,liu2021towards,C7SC04205A}
or linear scaling strategies 
\cite{saebo1993local,aquilante2006fast,li2009local,cole2010protein,wu2011linear,li2013linear,riplinger2013natural,li2014localization,li2005localized,li2017localization,ni2019fully}
when combined with efficient load-balancing schemes.
For example,the identification {of important interactions} of the SARS-CoV-2 spike protein at the QM level can be routinely implemented using fragment molecular orbitals (FMO) and molecular fractionation with conjugate caps (MFCC).\cite{kato2020intermolecular, akisawa2021interaction, akisawa2021fragment, fukuzawa2021special, wannipurage2022experiences, ma2023machine}
These advanced computational techniques provide a more detailed understanding of the electronic aspects of large molecules, {including} bio-pharmaceutical systems\cite{li2021perspective}.

As exascale supercomputing {progresses}, high-performance computing (HPC) is {playing} an increasingly crucial role in scientific calculations. {Gustafson's law asserts that the theoretical speed-up in parallel computing increases as resources increase.}\cite{gustafson1988reevaluating}
In quantum chemistry (QC) calculations, many algorithms can {be effectively parallelized. This includes} fragmentation calculations used in bio-pharmaceutical and solvent systems,\cite{chen2022carnot} as well as the high-throughput calculations {required for the} materials genome project and machine learning/artificial intelligence (ML/AI) tasks.
\cite{green2017fulfilling, de2019new, liu2020machine, suh2020evolving, pyzer2022accelerating}

Effective load balancing in scientific computing on HPC systems is crucial to ensure efficient resource utilization. {It involves} distributing sub-tasks across computing units to optimize response times and prevent overloading some nodes while leaving others idle. \cite{fujita2018development,ni2019analytical,abraham2020selected,wang2022low} Prior to load scheduling or distribution, the computational costs should be roughly predicted.
There are two {primary methods} used to predict job computational costs. {The first} assumes that similar tasks have similar costs, employing time series analysis and heuristic load balancing\cite{alexeev2012heuristic, gaussier2015improving}. {The second method} relies on computer system architecture, using machine learning (ML) with component performance data for cost prediction.\cite{helmy2015machine, shulga2016scheduling}
Very recently, several state-of-the-art parallel schemes have emerged. These include multi-level parallelization developed for large-scale QM calculations, where load-balancing can be effectively managed at each level, \cite{fedorov2023multi}and ML-assisted parallelization{, where ML/AI models provide more reliable predictions for computational costs.\cite{ma2023machine}}

In ML-assisted parallelization, static load balancing (SLB) can be pre-scheduled based on predictions of computational times for the subsystems to be calculated. 
\cite{ma2023machine} Several {methods} can be used {to predict} computational costs, such as the quantum ML approach proposed by Heinen and co-workers
\cite{heinen2020machine}, and the pre- and post-processing mechanism proposed by Wei and co-workers.
\cite{Jianwen2021predicting} In fact, strong generalization ML models inspired by the multiverse Ansatz, proposed by some of us
\cite{ma2021forecasting}, have already been introduced in previous ML-assisted parallelization work.
\cite{ma2023machine}
It is important to note that the precision of these predictions directly affects parallel efficiency. {While dynamic load balancing (DLB) can be used as a remedy to improve efficiency,} more reliable SLB schemes remain crucial for optimizing parallel efficiency. This work aims to explore the influence of fine descriptors on model predictions from our previous work, {thereby improving} the precision of predictions and, correspondingly, the parallel efficiency.

{However, it is important to recognize} that advanced parallel schemes typically achieve a high computational efficiency, {but they are not the sole} factor influencing the entire computational flow path. In distributed systems, certain nodes may encounter failures and become "straggler nodes" .
\cite{JMLR:v17:15-149}
The occurrence of straggler nodes can introduce unpredictable delays {while waiting for} feedback from these nodes, significantly affecting the overall efficiency of the distributed system.
\cite{pmlr-v70-tandon17a}

In recent years, researchers have ingeniously applied principles of coding theory to distributed computing, {giving rise} to the concept of coded distributed computing (CDC).
\cite{Li2016AFT}
CDC utilizes coding's flexibility to create redundant computations, {thereby reducing the} need for data exchanges between nodes. 
This type of approaches not only reduce communication time and load, but also mitigate the latency issues caused by straggler nodes.
Attia et al. proposed an almost optimal data shuffling coding scheme that achieves {a superior balance between storage and communication with fewer than five active} threads.
\cite{attia2019near}
To combat delays caused by straggler nodes in distributed computing, S. Kianidehkordi et al. implemented hierarchical coding in matrix multiplication{. This approach correlates} the task allocation of coded matrix multiplication with the geometric {challenge} of rectangular partitioning, which significantly enhances performance in computation, decoding, and communication times.
\cite{Kianidehkordi2021HierarchicalCM}
{Furthermore,} Sun et al. developed the hierarchical short-dot (HSD) computing scheme, {which employs hierarchical coding to} allow straggler nodes to handle fewer computing tasks. Coded computations also play a crucial role in secure and trustworthy distributed computing,{ bolstering system security and safeguarding user data privacy.}
\cite{wang2022optimal}

The {concept of} "fault tolerance" in distributed computing is a {pivotal element} of CDC{, enabling the correction of performance degradations caused} by lagging or disconnected nodes.
\cite{chen2005fault, fahim2021numerically, li2023fault}
{In} distributed scientific computing, {"fault tolerance" extends beyond managing such nodes to include the capacity for handling nodes that produce computational errors, while still ensuring} accurate overall results. {In communication systems, established coding methods like the Hamming code are utilized to ensure "fault tolerance" and enhance system stability.
\cite{hammingcodewiki}}
{Similarly, in distributed computation, various coding techniques, including the gradient coding scheme by Tandon et al., are employed to maintain "fault tolerance" throughout the computing process.
\cite{pmlr-v70-tandon17a}
We apply these principles to quantum chemical calculations to improve the stability of distributed cluster computing. 
For instance,} ML-assisted coded computations are employed in both the ground states and the excited states calculations, by MFCC approach and the re-normalized exciton model with time-dependent density functional theory (REM-TDDFT), respectively.\cite{ma2013calculating, wang_2022_Lowscaling}
{These methods allow} {the system to be divided into fragments, with energy-based or effective-Hamiltonian-based assembling can be used to recover }{the properties of the entire system.}

{The structure of the paper is organized as follows: Section 2 provides a concise overview of several key topics, including} 1) improved ML-assisted time predictions; 2) coded computing and ML-assisted coded scheduling optimization; 3) integration with the REM-TDDFT, and 4) implementations and workflow. Next, benchmark calculations of P38 protein, {solvent model with single and multiple excitable centres,} are presented in Section 3. Finally, we draw our conclusions in Section 4.

\section{\sffamily \Large Methodology and Implementation}

\subsection{\sffamily \large Improved ML-assisted time predictions}

In our previous work on predicting computational time, the input vector to the fully connected layer was a concatenation {that included the molecular structure feature vector alongside a singular value representing the total number of basis sets.
\cite{ma2021forecasting}
Clearly, using just one number to encapsulate the information about basis sets is overly simplistic and insufficient for detailed descriptions. 
To achieve more precise predictions, incorporating more detailed descriptors of basis sets into the ML model is essential. Indeed, similar approaches have already been implemented successfully} in various quantum tensor learning models.
\cite{schutt2019unifying, gastegger2020deep, unke2021se, li2022deep, gong2023general, li2023deep}

\begin{figure}[htbp]
\centering
\includegraphics[scale=1.00, bb=50 0 400 280]{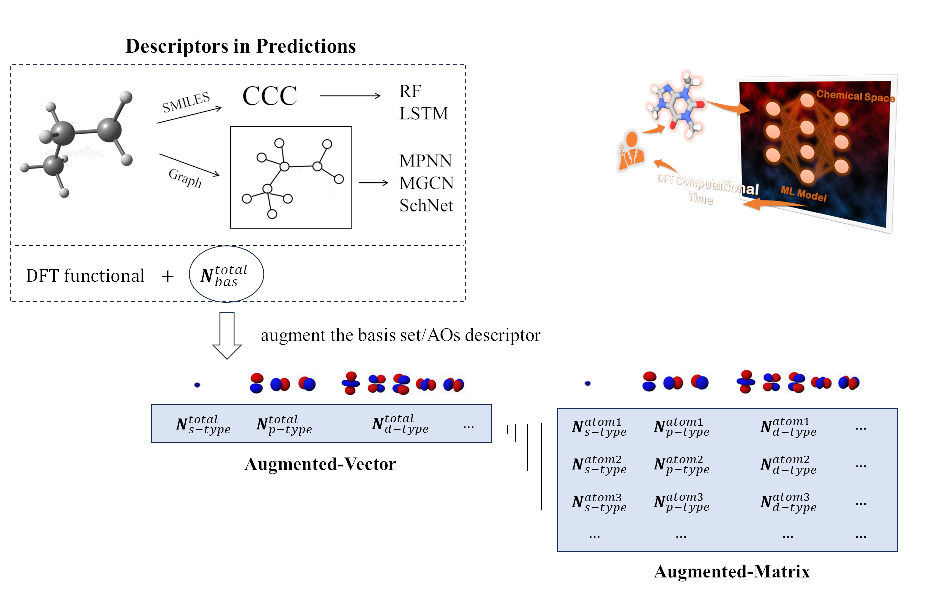}
\caption{The idea for the improvement of basis set/AOs descriptors in predicting models.}
\label{improve_idea}
\end{figure}

According to the linear combination of atomic orbitals (LCAO) ansatz, the basis sets {encompass all atomic orbitals (AOs) represented in the expansion.
\cite{LCAOwiki}
Once these basis sets are defined, the total number of AOs involved in the calculations can be readily determined, allowing for an initial count of basis functions prior to practical computations. This method was employed in our previous works for training machine learning models, although the approach was relatively rudimentary
\cite{ma2021forecasting, ma2023machine},
denoted as "original" in this work.}
Herein, two progressive descriptors for describing the basis functions are introduced as shown in Fig.\ref{improve_idea}. {The first descriptor, termed "augmented-vector" or "aug-V" for short, captures} the subtotal numbers of different types of AOs (such as $s$, $p$, $d$ orbitals) separately. This augmented-vector is then integrated with the molecular structure feature vector in the fully connected layer of the time prediction ML model. {The second descriptor, called "augmented-vector" or "aug-V", records} the subtotal numbers of different types of AOs for each individual atom. Each atom's AOs are represented as row vectors in this matrix and are included in the nodes' features within the graph. {Consequently, the input to the graph includes vectors of} these subtotal numbers for individual atoms in a molecule.

\begin{figure}[htbp]
\centering
\includegraphics[scale=1.00, bb=40 0 400 185]{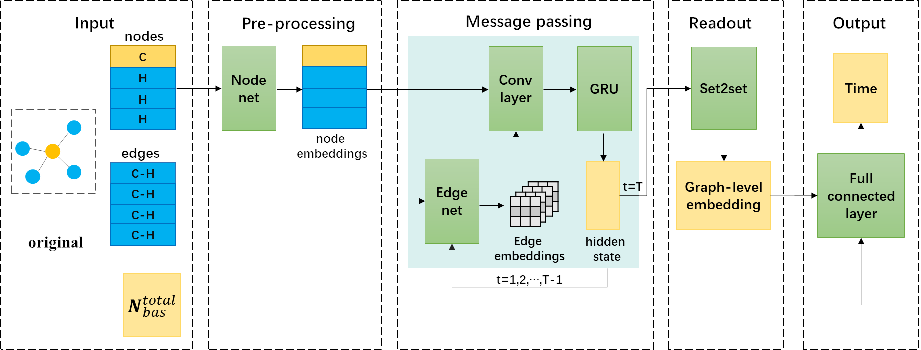}
\includegraphics[scale=1.00, bb=40 0 400 185]{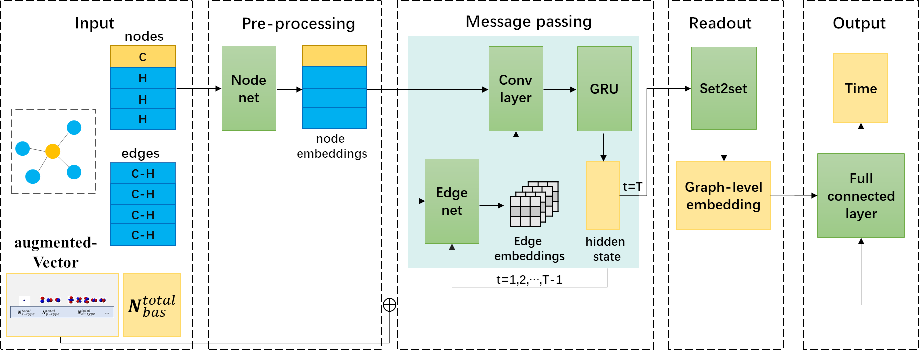}
\includegraphics[scale=0.9815, bb=50 0 400 185]{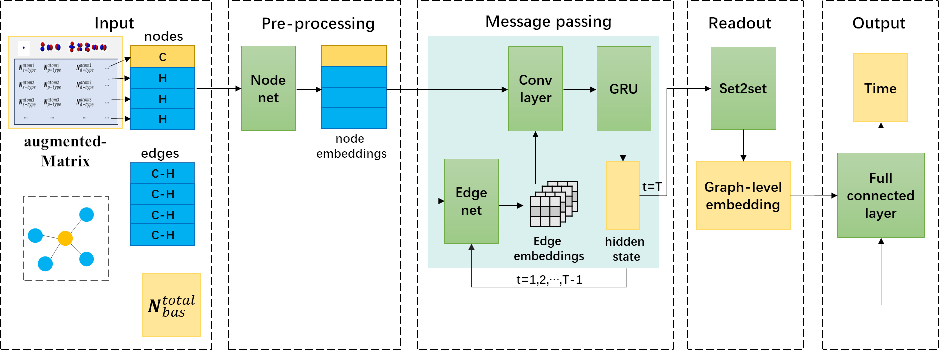}
\caption{The process structure of variant MPNN models using original (\textit{upper}), augmented-vector (\textit{middle}), and augmented-matrix (\textit{bottom}) descriptors, separately. Other graph-based models, e.g. MGCN, can share the similar process structure when using the improved descriptors.}
\label{structure_MPNN_improve}
\end{figure}

In the "aug-V" variant, the sub-total numbers of different type of AOs for a molecule can be made into a vector in a batch, i.e. 
\begin{equation*}
  vector = [n_s, n_p, n_d, n_f, n_g, \cdots ].
\end{equation*}
The topological structure of the "aug-V" variant in working models is {depicted} in Fig.\ref{structure_MPNN_improve}, where the vector is pre-embedded in a sample message-passing neural network (MPNN) model.
\cite{gilmer2017neural}
In the "aug-M" variant, the total numbers of different types of atomic orbitals for each atom can be organized into a matrix in a batch, i.e.
\begin{equation*}
  matrix =
  \begin{bmatrix}
  n_{s}^{1} & n_{p}^{1} & n_{d}^{1} & n_{f}^{1} & n_{g}^{1} & \cdots \\
  n_{s}^{2} & n_{p}^{2} & n_{d}^{2} & n_{f}^{2} & n_{g}^{2} & \cdots \\
   \cdots &  \cdots &  \cdots&  \cdots & \cdots  & \cdots \\
  n_{s}^{N} & n_{p}^{N} & n_{d}^{N} & n_{f}^{N} & n_{g}^{N} & \cdots \\
\end{bmatrix}.
\end{equation*}
The "aug-M" variant differs in that the sub-total numbers of different types of atomic orbitals are not integrated into the array of basis functions numbers but are directly combined with the features of the nodes in the molecular topology diagram. Therefore, this variant can only be used in graph-based models, e.g., MPNN and multi-level graph convolutional neural network (MGCN) 
\cite{lu2019molecular}, which accept molecular topology diagrams as input. For long short-term memory (LSTM) 
\cite{graves2005bidirectional} or random forest models, which coordinate the simplified molecular-input line-entry system (SMILES) code, cannot be adapted within this framework. The topological structure of the "aug-M" variant in working models is illustrated in Fig.\ref{structure_MPNN_improve} with embedding in a sample MPNN model.

\subsection{\sffamily \large Coded computing and coded scheduling optimization}

\subsubsection{\sffamily Coded computing}

In general, conventional distributed computing tasks can be simplified as the model shown in Fig.\ref{Fig_gradient_coding-1}: 
individual nodes ($w_i$) are assigned different computing tasks ($D_i$), and nodes compute properties ($g_i$) of the overall system, which are then aggregated to obtain the overall properties of the system ($g$).
It is important to note that, in the current scenario, the final aggregation operation is summed {directly by receiving} {results from various computing nodes to the master node. Notice that there is no fault tolerance in this case, meaning any abnormality in any node/task can lead to unreliable results.
For introducing fault tolerance, the following questions need addressing: 1) how to distribute tasks with redundancy to nodes; 2) how to aggregate the results at the master node in the presence of redundant outcomes.}

\tikzset{every picture/.style={line width=0.75pt}} 
\begin{figure}[htbp]
  \centering
  \tikzset{every picture/.style={line width=0.75pt}} 
  \subfigure[Naive synchronous gradient descent]{
  \begin{tikzpicture}[x=0.75pt,y=0.75pt,yscale=-1,xscale=1]
  
  \draw   (230.4,100.2) -- (270,100.2) -- (270,139.67) -- (230.4,139.67) -- cycle ;
  \draw   (160.4,191) -- (200,191) -- (200,230.47) -- (160.4,230.47) -- cycle ;
  \draw   (230.4,191) -- (270,191) -- (270,230.47) -- (230.4,230.47) -- cycle ;
  \draw   (300,190.6) -- (339.6,190.6) -- (339.6,230.07) -- (300,230.07) -- cycle ;
  \draw    (200,191) -- (229.38,141.39) ;
  \draw [shift={(230.4,139.67)}, rotate = 120.63] [color={rgb, 255:red, 0; green, 0; blue, 0 }  ][line width=0.75]    (10.93,-3.29) .. controls (6.95,-1.4) and (3.31,-0.3) .. (0,0) .. controls (3.31,0.3) and (6.95,1.4) .. (10.93,3.29)   ;
  \draw    (250,190.47) -- (250,142.47) ;
  \draw [shift={(250,140.47)}, rotate = 90] [color={rgb, 255:red, 0; green, 0; blue, 0 }  ][line width=0.75]    (10.93,-3.29) .. controls (6.95,-1.4) and (3.31,-0.3) .. (0,0) .. controls (3.31,0.3) and (6.95,1.4) .. (10.93,3.29)   ;
  \draw    (300,190.6) -- (271.02,141.39) ;
  \draw [shift={(270,139.67)}, rotate = 59.5] [color={rgb, 255:red, 0; green, 0; blue, 0 }  ][line width=0.75]    (10.93,-3.29) .. controls (6.95,-1.4) and (3.31,-0.3) .. (0,0) .. controls (3.31,0.3) and (6.95,1.4) .. (10.93,3.29)   ;
  
  \draw (241,111.8) node [anchor=north west][inner sep=0.75pt]  [font=\fontsize{1.2em}{1.44em}\selectfont] [align=left] {W};
  \draw (168.2,202.6) node [anchor=north west][inner sep=0.75pt]  [font=\fontsize{1.2em}{1.44em}\selectfont] [align=left] {D1};
  \draw (238.2,203) node [anchor=north west][inner sep=0.75pt]  [font=\fontsize{1.2em}{1.44em}\selectfont] [align=left] {D2};
  \draw (307.4,203) node [anchor=north west][inner sep=0.75pt]  [font=\fontsize{1.2em}{1.44em}\selectfont] [align=left] {D3};
  \draw (167.33,233.47) node [anchor=north west][inner sep=0.75pt]   [align=left] {w1};
  \draw (239.33,233.07) node [anchor=north west][inner sep=0.75pt]   [align=left] {w2};
  \draw (309.73,232.67) node [anchor=north west][inner sep=0.75pt]   [align=left] {w3};
  \draw (190.13,161.27) node [anchor=north west][inner sep=0.75pt]    {$g_{1}$};
  \draw (232.73,161.07) node [anchor=north west][inner sep=0.75pt]    {$g_{2}$};
  \draw (271.93,160.47) node [anchor=north west][inner sep=0.75pt]    {$g_{3}$};
  \draw (215.73,83.27) node [anchor=north west][inner sep=0.75pt]  [font=\small]  {$g_{1} +g_{2} +g_{3}$};

  \end{tikzpicture}
  \label{Fig_gradient_coding-1}
  }
  \quad
  \subfigure[Gradient coding: The vector g1 + g2 + g3 is in the span of any two out of the vectors g1/2 + g2, g2 - g3 and g1/2 + g3.]{    

  \begin{tikzpicture}[x=0.75pt,y=0.75pt,yscale=-1,xscale=1]
  
  \draw   (250.4,120.2) -- (290,120.2) -- (290,159.67) -- (250.4,159.67) -- cycle ;
  \draw   (250.37,210.1) -- (289.97,210.1) -- (289.97,249.57) -- (250.37,249.57) -- cycle ;
  \draw   (320,210.6) -- (359.6,210.6) -- (359.6,250.07) -- (320,250.07) -- cycle ;
  \draw    (270,210.47) -- (270,162.47) ;
  \draw [shift={(270,160.47)}, rotate = 90] [color={rgb, 255:red, 0; green, 0; blue, 0 }  ][line width=0.75]    (10.93,-3.29) .. controls (6.95,-1.4) and (3.31,-0.3) .. (0,0) .. controls (3.31,0.3) and (6.95,1.4) .. (10.93,3.29)   ;
  \draw    (320,210.6) -- (291.02,161.39) ;
  \draw [shift={(290,159.67)}, rotate = 59.5] [color={rgb, 255:red, 0; green, 0; blue, 0 }  ][line width=0.75]    (10.93,-3.29) .. controls (6.95,-1.4) and (3.31,-0.3) .. (0,0) .. controls (3.31,0.3) and (6.95,1.4) .. (10.93,3.29)   ;
  \draw    (250.33,229.83) -- (290,229.83) ;
  \draw    (319.97,230.33) -- (359.63,230.33) ;
  \draw   (180.07,210.93) -- (219.67,210.93) -- (219.67,250.4) -- (180.07,250.4) -- cycle ;
  \draw    (180,229.77) -- (219.67,229.77) ;
  \draw    (219.67,210.93) -- (249.37,161.38) ;
  \draw [shift={(250.4,159.67)}, rotate = 120.94] [color={rgb, 255:red, 0; green, 0; blue, 0 }  ][line width=0.75]    (10.93,-3.29) .. controls (6.95,-1.4) and (3.31,-0.3) .. (0,0) .. controls (3.31,0.3) and (6.95,1.4) .. (10.93,3.29)   ;
  
  \draw (261,131.8) node [anchor=north west][inner sep=0.75pt]  [font=\fontsize{1.2em}{1.44em}\selectfont] [align=left] {W};
  \draw (189.33,253.13) node [anchor=north west][inner sep=0.75pt]   [align=left] {w1};
  \draw (259.33,253.07) node [anchor=north west][inner sep=0.75pt]   [align=left] {w2};
  \draw (329.73,252.67) node [anchor=north west][inner sep=0.75pt]   [align=left] {w3};
  \draw (174.93,187.67) node [anchor=north west][inner sep=0.75pt]  [font=\scriptsize]  {$g_{1} /2+g_{2}$};
  \draw (234.67,187.4) node [anchor=north west][inner sep=0.75pt]  [font=\scriptsize]  {$g_{2} -g_{3}$};
  \draw (315.07,187.47) node [anchor=north west][inner sep=0.75pt]  [font=\scriptsize]  {$g_{1} /2+g_{3}$};
  \draw (235.73,103.27) node [anchor=north west][inner sep=0.75pt]  [font=\small]  {$g_{1} +g_{2} +g_{3}$};
  \draw (189.89,212.78) node [anchor=north west][inner sep=0.75pt]   [align=left] {D1};
  \draw (329.56,232.44) node [anchor=north west][inner sep=0.75pt]   [align=left] {D1};
  \draw (189.22,232.78) node [anchor=north west][inner sep=0.75pt]   [align=left] {D2};
  \draw (259.56,212.11) node [anchor=north west][inner sep=0.75pt]   [align=left] {D2};
  \draw (259.89,231.78) node [anchor=north west][inner sep=0.75pt]   [align=left] {D3};
  \draw (329.56,212.78) node [anchor=north west][inner sep=0.75pt]   [align=left] {D3};

  \end{tikzpicture}
  \label{Fig_gradient_coding-2}
  }

  \caption{The idea of gradient coding}\label{Fig_gradient_coding}
  \end{figure}
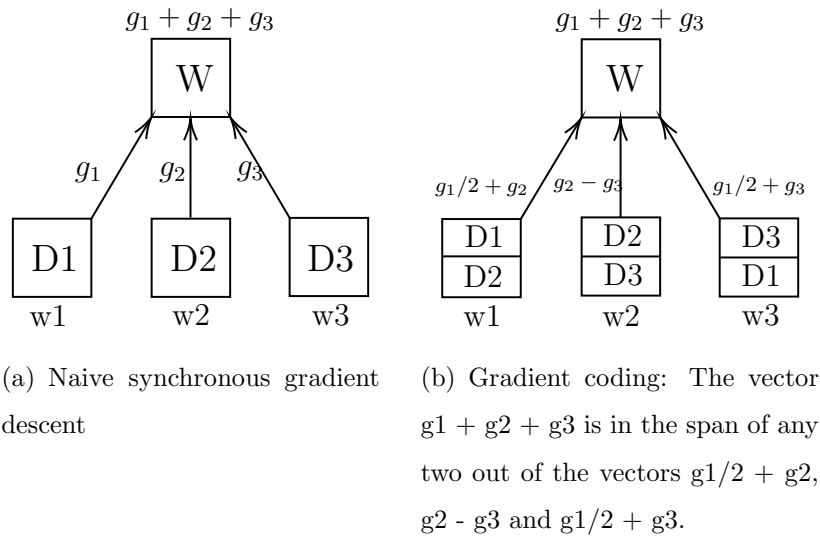

{In the context of coded computing proposed by Tandon et al. 
\cite{pmlr-v70-tandon17a}, we illustrate the overall process using gradient coding as an example to introduce fault tolerance.}
In the framework of {coded} computation as shown in Fig.\ref{Fig_gradient_coding-2}, each node is assigned an additional computing task ($D_{j, j\neq i}$), and the results of the tasks undertaken by nodes are multiplied by different coefficients, {involving coding and decoding procedures.}
{This setup ensures that} even if one node produces an incorrect result or drops out, the correct overall properties can still be obtained through the calculated results of other nodes.
Tandon et al. pointed out that {this process can be achieved by constructing two special matrices (operations), namely} decoding ($A$) and encoding ($B$).
For the scenario shown in Fig.\ref{Fig_gradient_coding-2}, the matrices $A$ and $B$ can be expressed as {follows:}

\begin{equation*}
A = \begin{bmatrix}
  0 & 1 & 2 \\
  1 & 0 & 1 \\
  2 & -1 & 0 \\
\end{bmatrix},   
 B = \begin{bmatrix}
  1/2 & 1 & 0 \\
  0 & 1 & -1 \\
  1/2 & 0 & 1 \\
\end{bmatrix}. 
\end{equation*}

{Herein, each row of the matrix $B$ (encoding matrix) corresponds to a computing node, {where} all elements in the same column represent coefficients for a specific task and its copies. 
If the coefficient \textit{B}(\textit{i,j}) is 0, it indicates that the \textit{j}th task will not be executed by the \textit{i}th node. 
On the other hand, the matrix $A$ serves as the decoding matrix, with each row representing a potential successful solution when multiplied with the matrix $B$.}

{The final result can be reconstructed using the matrix $C \textbf{g}$ ($C = A \times B$) {through the operation} $\sum_j C(i,j) \cdot g_j$, i.e. 
\begin{equation*}
 C \textbf{g} = (A \times B) \textbf{g} = \begin{bmatrix}
  1 & 1 & 1 \\
  1 & 1 & 1 \\
  1 & 1 & 1 \\
\end{bmatrix}
\begin{bmatrix}
 g_1 \\
 g_2 \\
 g_3 \\
\end{bmatrix}
= \begin{bmatrix}
 g_1 + g_2 + g_3 \\
 g_1 + g_2 + g_3 \\
 g_1 + g_2 + g_3 \\
\end{bmatrix},
\end{equation*}
and the row dimension of matrix $C$ reflects the corresponding redundant {levels}. 
If no abnormalities occur, three identical results can be decoded.
For example, if the \textit{1}st node (w1) lags behind or produces an incorrect result, the $C \textbf{g}$ vector will become 
$[ g_1 + g_2 + g_3 ,  \text{err},   \text{err} ]^T$, and there is still one correct result that can be decoded. 
{The divergent} decoded results in the $C \textbf{g}$ array can {indicate} either broken node(s) (if elements are missing in $C \textbf{g}$) or convergence issues (if deviations are observed in $C \textbf{g}$) in quantum chemical calculations, respectively.
}

\subsubsection{\sffamily Coded scheduling optimization}
To improve the stability of distributed quantum chemical calculations, the encoding-decoding process is incorporated into the ML-assisted scheduling optimization.
The pseudo-codes presented in Algorithm \textbf{1} and \textbf{2} can be utilized for generating the {corresponding encoding ($B$) and decoding ($A$) matrices}, respectively.

\begin{algorithm} 
  \caption{Calculate {the encoding} matrix $B$} 
    \begin{algorithmic}
    \REQUIRE $ n,s(<n)$    {$ \quad \vartriangleright n : total \ nodes, \ s : stragger \ node(s) $}
    \ENSURE $B \in R^{n\times n}$ with $ (s + 1)$ non-zeros in each row
    \STATE $H = $randn$(s,n)$ 
    \STATE $H(:,n) = -$sum$(H(:,1:n-1), 2)$; 
    \STATE $B = $ zeros$(n)$; 
    \FOR{$i = 1: n$} 
    \STATE $j =$ mod$(i - 1:s + i - 1, n) + 1$ ;
    \STATE $B(i,j) = [1; -H(:,j(2:s+1)) \backslash H(:,j(1))]$; 
    \ENDFOR
    \end{algorithmic} 
\end{algorithm}

\begin{algorithm} 
  \caption{Calculate {the decoding} matrix $A$} 
  \begin{algorithmic}
  \REQUIRE matrix$B$ 
  \ENSURE matrix$A$ 
  \FOR{$I\subseteq [n]$, $|I| = (n - s)$} 
  \STATE $a =$ zeros$(1,n)$ ;
  \STATE $x =$ ones$(1,n)/B(I,:)$; 
  \STATE $a(I) = x$;
  \STATE $A = [A;a]$;
  \ENDFOR
  \end{algorithmic} 
\end{algorithm}

{However, it is common for only certain specific regions in molecular systems to encounter instability during calculations, such as binding domains or excited states. As encoding calculations can be computationally demanding, not all tasks need to be encoded.} 
Thus, in the subsequent process, the distributed tasks can be {divided into two categories}: encoded tasks and non-encoded tasks. 
These two task types are {handled sequentially}, i.e. 1) specify the encoded tasks and their computing groups 2) {perform} scheduling optimization for all tasks.

\subsection{\sffamily \Large REM-TDDFT approach with coded optimization}
In our very recent work, the coded framework for {computing the} ground state binding energy in protein-ligand systems is preliminarily {discussed} \cite{li2024coded}. 
Herein, we {integrate} ML-assisted coded computing with the real-space re-normalization approach, specifically REM-TDDFT, {i.e. REM-TDDFT, aiming for the automatic calculation of excited states.}

\begin{figure}[htbp]
  \centering
    \includegraphics[scale=0.085,bb=0 0 4000 1200]{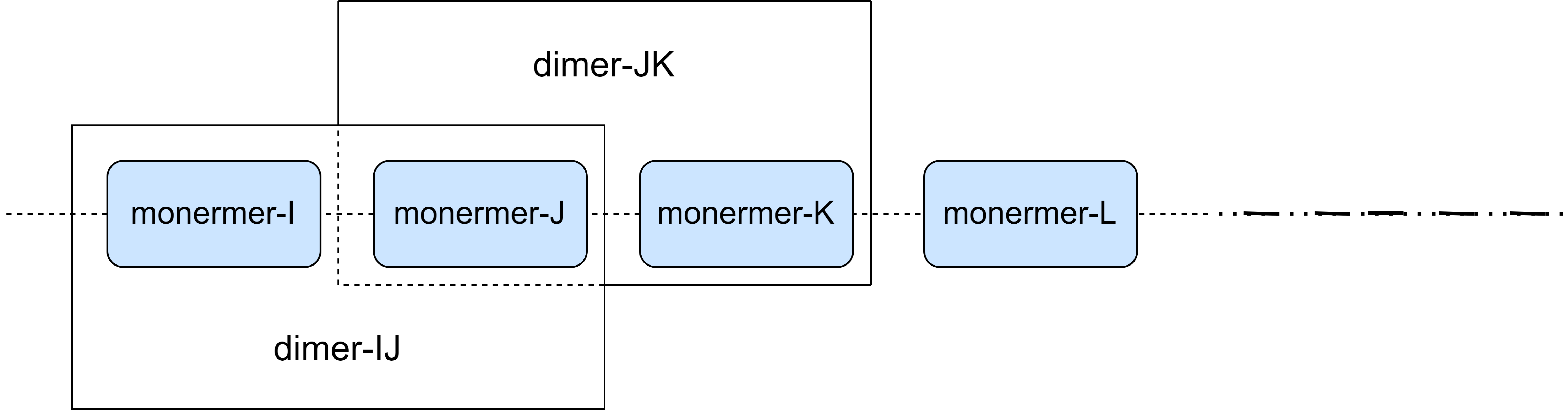}
    \caption{Partition of the whole system into various blocks.}
    \label{partition_of_system}
\end{figure}

The REM is a type of real-space or fragment-based method. 
In REM, the whole system can be divided into many sub-systems or blocks (usually tens or hundreds), as illustrated in Fig.\ref{partition_of_system}.
Here the $I, J, K $ and $L$ are the monomers, 
and additionally, the adjacent monomers form the dimers.
{In REM anastz, one assumes that the wave function of excited states of the whole system ($|\Psi_{REM}\rangle$) can be treated as the linear combination of local excited states in the sub-systems or blocks,\cite{ma2012new, ma2013calculating} i.e. 
\begin{equation}
 |\Psi_{REM}\rangle =  \sum\limits_{I=1}^{N} C_I [|\psi^*_I\rangle\prod\limits_{J\neq I}|\psi_J^0 \rangle]
\label{eqProj}
\end{equation}
in which the superscript $*$, $0$ represent the the excited or ground state, respectively, for a specific sub-system. 
Once the model space (wave function space) is confirmed, the effective Hamiltonian can be constructed using the following expressions,
\begin{equation}
H^{eff} = \sum\limits_{I = 1}^N H^{eff}_{I} + \sum\limits_{I > J} H^{eff}_{I,J} + \cdots
\label{eqREMHami}
\end{equation}
where the one-body terms ($H^{eff}_{I}$, normally treated as $\Delta E_{I^{*}}$) and the $N$-body interaction terms can be used to construct the $H^{eff}$. The Eq.(\ref{eqREMHami}) can be rewritten in the matrix form, i.e. 
\begin{equation}
\begin{aligned}
|A^*B^0C^0\cdots⟩   \\
|A^0B^*C^0\cdots⟩   \\
|A^0B^0C^*\cdots⟩   \\
\cdots \qquad \\
\end{aligned}
\begin{bmatrix}
\Delta E_{A^*} & h_{AB} & h_{AC} & \cdots \\
h_{AB} & \Delta E_{B^*} & h_{BC} & \cdots \\
h_{AC} & h_{BC} & \Delta E_{C^*} & \cdots \\
\vdots & \vdots & \vdots         & \ddots \\
\end{bmatrix}
\label{eqREMHamiM}
\end{equation}
where the $\Delta E_{I^{*}}$ is the local excitation variation on block $I$ in the presence of other blocks in their ground state, and $h_{IJ}$ is the excitation transfer value (i.e. hopping integral) between block $I$ and $J$.
The $\Delta E_{I^{*}}$ and $h_{IJ}$ can be obtained from solving exactly the many-block problems via the Bloch's effective Hamiltonian theory. 
Take the evaluation of $\Delta E_{A^{*}}$, $\Delta E_{B^{*}}$, and $h_{AB}$ as example, the following step-by-step instruction can be used :  
\begin{itemize} 
\item Obtain the ground and 1st excited states for monomer-$A$, e.g. $H_{A}|\psi^0_{A}\rangle = E_{A}^0|\psi^0_{A}\rangle$ and $H_{A}|\psi^*_{A}\rangle  = E_{A}^*|\psi^*_{A}\rangle$; same for monomer-$B$. 
\item Construct the block product basis $|\psi^*_{A}\psi^0_{B}\rangle$ and $|\psi^0_{A}\psi^*_{B}\rangle$, and the projector in this model space, i.e. $P_{AB}^* =|\psi^*_A\psi_B^0\rangle \langle \psi_A^*\psi_B^0| + |\psi^0_A\psi_B^*\rangle \langle \psi_A^0\psi_B^*|$
\item Obtain the 1st and 2nd excited states of dimer-$AB$, e.g. $H_{AB}|\psi^*_{AB}\rangle  = E_{AB}^*|\psi^*_{AB}\rangle$, and $H_{AB}|\psi^{*'}_{AB}\rangle = E_{AB}^{*'}|\psi^{*'}_{AB}\rangle$, respectively. 
\item Project the excited states of dimer-$AB$ into model space, e.g. $P_{AB}^*|\psi^*_{AB}\rangle = a_{*0}|\psi^*_{A}\psi_{B}^0\rangle +  a_{0*}|\psi^0_{A}\psi_{B}^*\rangle$, and $P_{AB}^{*'}|\psi^{*'}_{AB}\rangle = a_{*'0}|\psi^{*}_{A}\psi_{B}^0\rangle +  a_{0*'}|\psi^0_{A}\psi_{B}^{*}\rangle$; Orthonormalizing the coefficients of projected vectors as $b_{*0}$, $b_{0*}$, $b_{*'0}$, and $b_{0*'}$, rewrite them as $2\times 2$ matrix $B$.
\item Obtain the $2\times 2$ Bloch's effective Hamiltonian matrix $H_{AB}^{eff} = B * E_{AB} * B^{\dagger}$, in which $E_{AB}^{eff}$ is the diagonal matrix with $E_{AB}^*$ and $E_{AB}^{*'}$.
\item Obtain the hopping integral $h_{AB}$ (equate to $H_{AB}^{eff}[1,2]$), local excitation variation $\Delta E_{A^{*}}$ (equate to $H_{AB}^{eff}[1,1] - E_{A}^* - E_{B}^0$), and $\Delta E_{B^{*}}$ (equate to $H_{AB}^{eff}[2,2] - E_{A}^0 - E_{B}^*$), separately.
\end{itemize}
}

When {combining} ML-assisted coded computing with REM-TDDFT, one can {start by assigning the potential excited monomers. Subsequently, the excitable dimers can be identified} by considering interacting monomers within a given distance, such as 4 \AA. {Based} on a specific excitable monomer, {all dimers containing this monomer, along with the other monomers in those dimers, can be grouped for coded computing.} 
All sub-tasks (monomers and dimers) in a group can be encoded {using the matrices $A$ and $B$, utilizing Algorithm \textbf{1} and \textbf{2}.} After that, the coded and non-coded sub-tasks can be {merged, and the scheduling can be optimized using a greedy algorithm based} on the predicted timings for the DFT/TDDFT calculations.

\subsection{\sffamily \Large Implementation and workflow}

The {\sc Fcst\_sys} package \cite{ma2021forecasting} is adapted for the augmented descriptor, {enabling more accurate predictions using improved ML models}. At the meantime, simple {\sc Matlab} or {\sc Python} script can be {employed} to generate the {encoding} and decoding matrices, which only involving the part/desired sub-tasks to be computed in the forthcoming distributed calculations. 
After encoding the desired sub-tasks, {each} encoded task-row, {representing sub-tasks in the $B$ matrix}, is treated as a single task.
The predicted computational costs are the summation of individual sub-tasks in the task-row. 

{Once the predicted computational costs for all sub-tasks, both coded and non-coded, are obtained}, planning algorithms like the greedy algorithm can be {utilized to derive} the proposed SLB scheduling. 
{The scheduling is implemented using a {\sc Python} script, which provides the recommended scheduling for distributed HPC calculations.} 
The {\sc ParaEngine} package, \cite{ma2023machine} {originally} developed as a demonstrative engine for ML-assisted quantum chemical calculations, can be {repurposed} as the coded quantum chemical engine with {enhanced} SLB scheduling. 
{Upon completion of all calculations}, the REM-TDDFT approach can be {employed to assess} the excited states of the whole systems.

\begin{figure} [!htbp]
\centering
	\includegraphics[scale=1.15,bb=05 0 550 210]{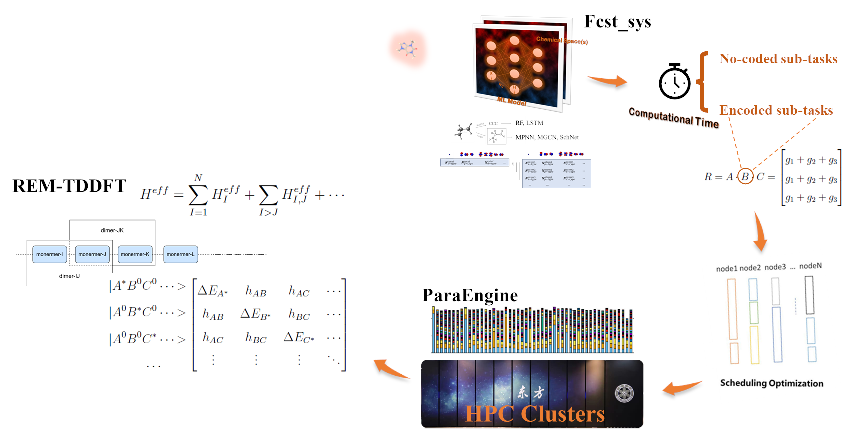}
	\caption{Workflow of the coded quantum chemical calculations with improved ML models. } \label{fig_workflow}
\end{figure}

The {entire} workflow is {depicted} in Fig.\ref{fig_workflow} for a visual illustration.

\section{\sffamily \Large RESULTS}

\subsection{\sffamily \Large Term-wise assessments}

\subsubsection{\sffamily \large Improved ML models}

The "original," "aug-V," and "aug-M" ML models, introduced in sub-section 2.1, {are evaluated using the same reference molecular suits that from our previous work. \cite{ma2021forecasting, Fcst_github}. 
The molecules in the reference suits are sampled from the {\sc DrugBank} database \cite{wishart2008drugbank} and {categorized} into different sub-suit in the order of molecular weight (e.g. small: 1-200; medium: 200-400; large: $>$ 400).
The calculated results and timing data are {generated using {\sc Gaussian09} with} the B3LYP functional and {various} basis sets, i.e. Pople's 6-31g, 6-31g*, and 6-31+g*, separately.}

\begin{figure}[htbp]
    \centering
    \subfigure[Different basis sets]{
    \begin{minipage}[t]{0.475\linewidth}
    \begin{overpic}[scale=0.50,bb=0 0 400 320]{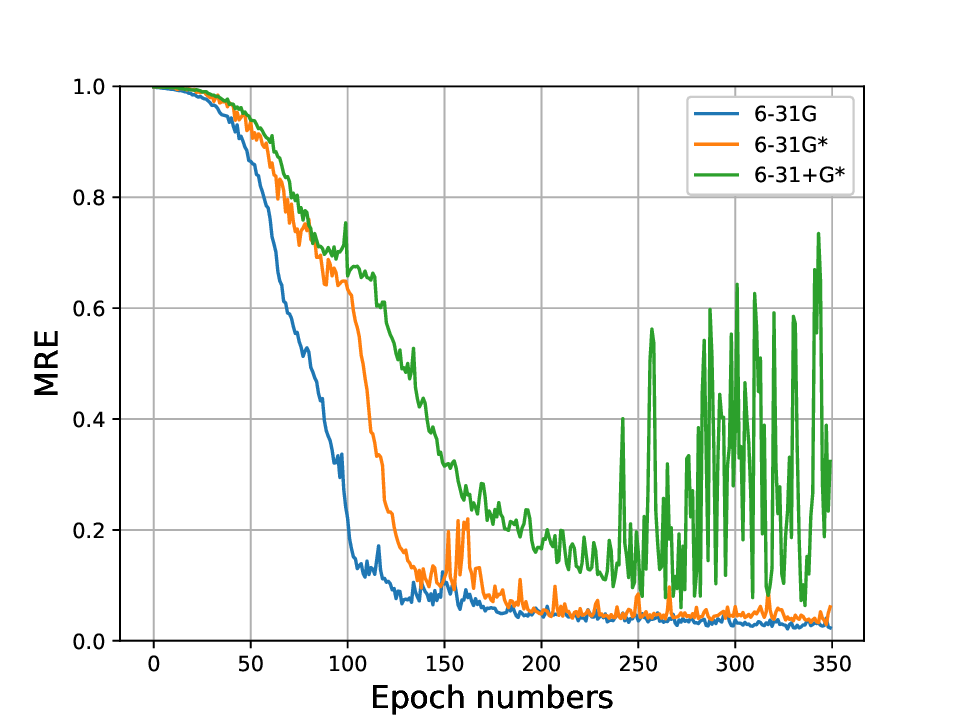}
      \put(63,54){original}
    \end{overpic}
    \begin{overpic}[scale=0.50,bb=0 0 400 310]{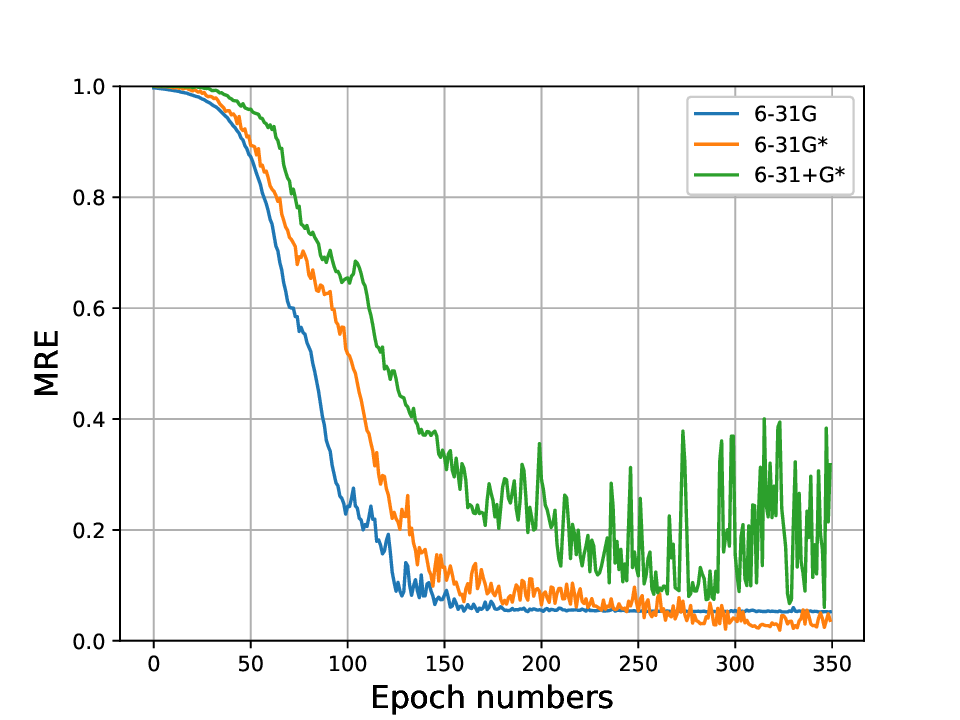}
      \put(63,54){aug-V}
    \end{overpic}
    \end{minipage}\label{Fig_LSTM-1}
    }
    \subfigure[Different molecular weights]{
    \begin{minipage}[t]{0.475\linewidth}
    \begin{overpic}[scale=0.50,bb=0 0 400 320]{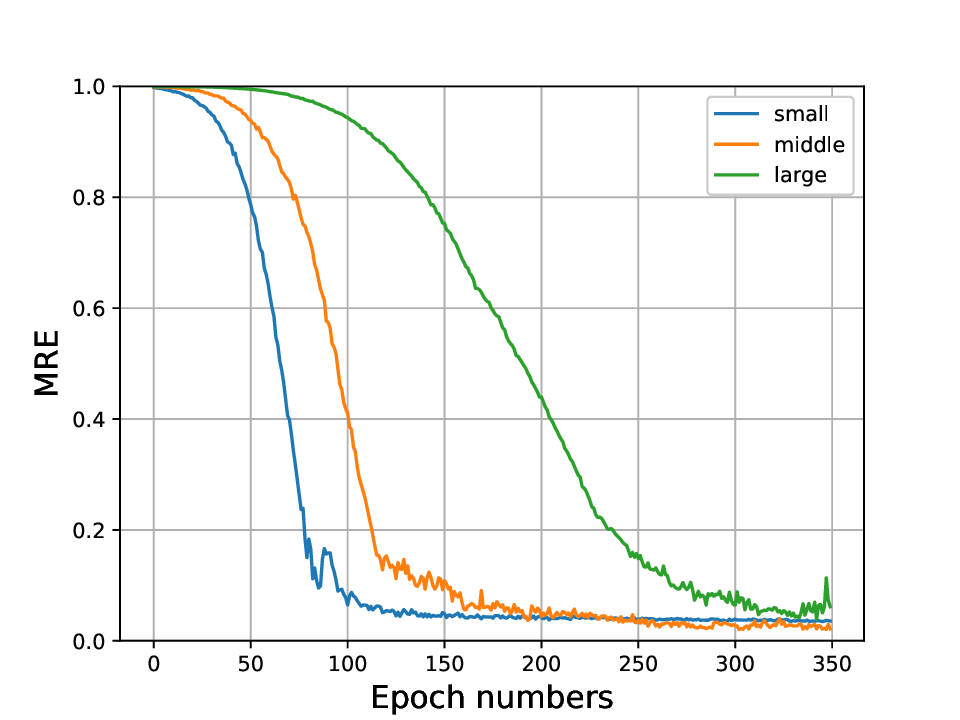}
      \put(68,54){original}
    \end{overpic}
    \begin{overpic}[scale=0.50,bb=0 0 400 310]{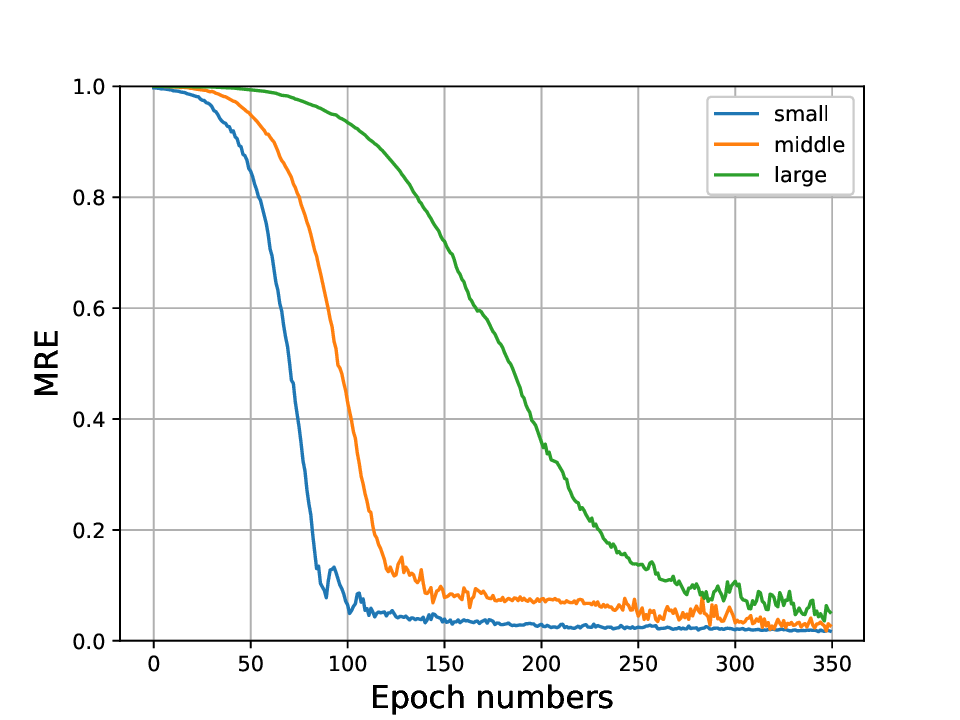}
      \put(68,54){aug-V}
    \end{overpic}
    \end{minipage}\label{Fig_LSTM-2}
    }
    \caption{The calculated MRE in the training process with "original" (upper) and "aug-V" (bottom) bi-LSTM models, respectively.}
    \label{Fig_LSTM-MRE}
    \end{figure}

\begin{figure}[htbp]
      \centering
      \subfigure[Different basis sets]{
      \begin{minipage}[t]{0.475\linewidth}
      \begin{overpic}[scale=0.50,bb=0 0 400 320]{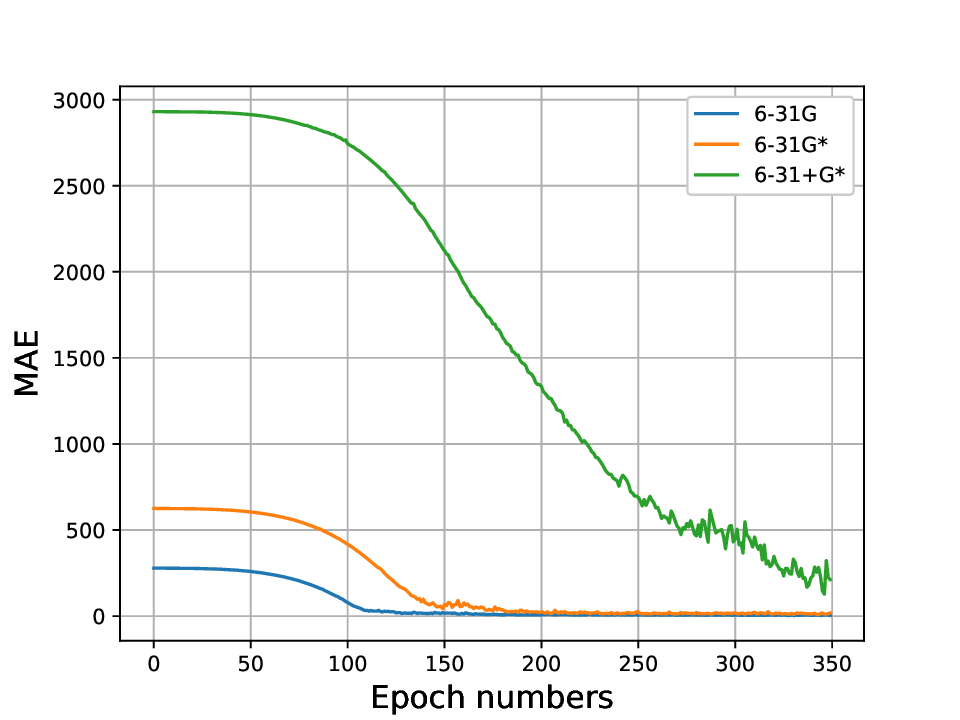}
        \put(63,54){original}
      \end{overpic}
      \begin{overpic}[scale=0.50,bb=0 0 400 310]{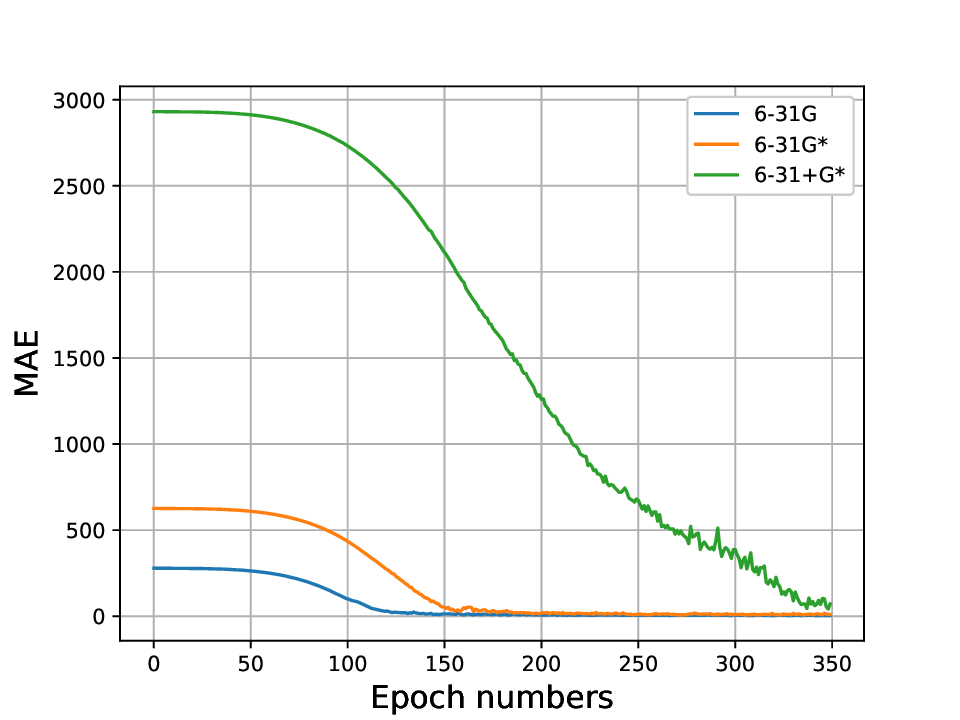}
        \put(63,54){aug-V}
      \end{overpic}
      \end{minipage}\label{Fig_LSTM-3}
      }
      \subfigure[Different molecular weights]{
      \begin{minipage}[t]{0.475\linewidth}
      \begin{overpic}[scale=0.50,bb=0 0 400 320]{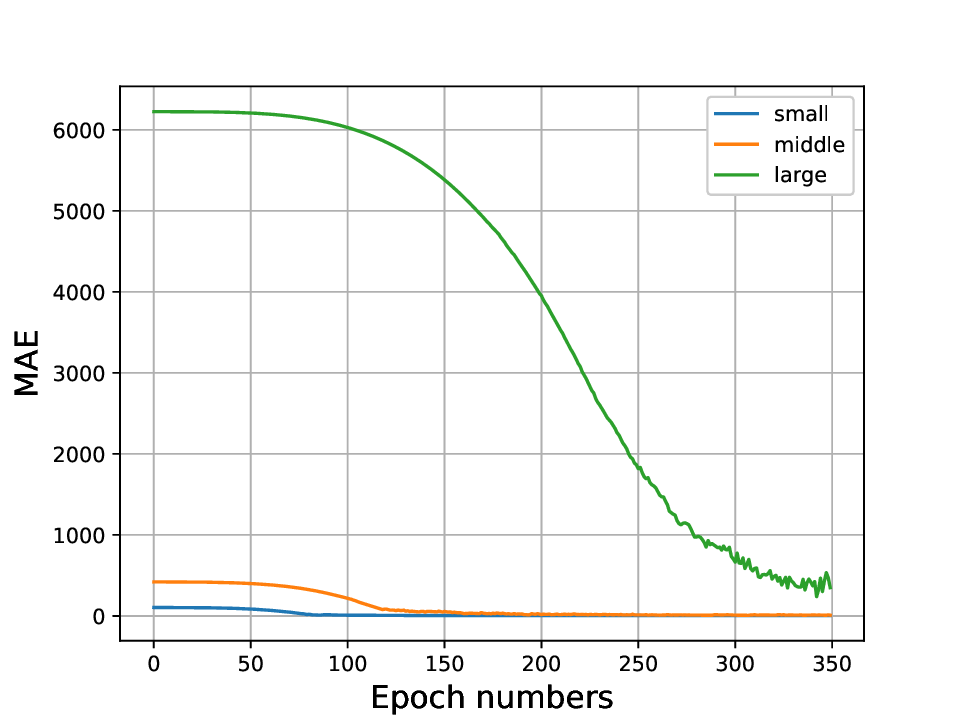}
        \put(68,54){original}
      \end{overpic}
      \begin{overpic}[scale=0.50,bb=0 0 400 310]{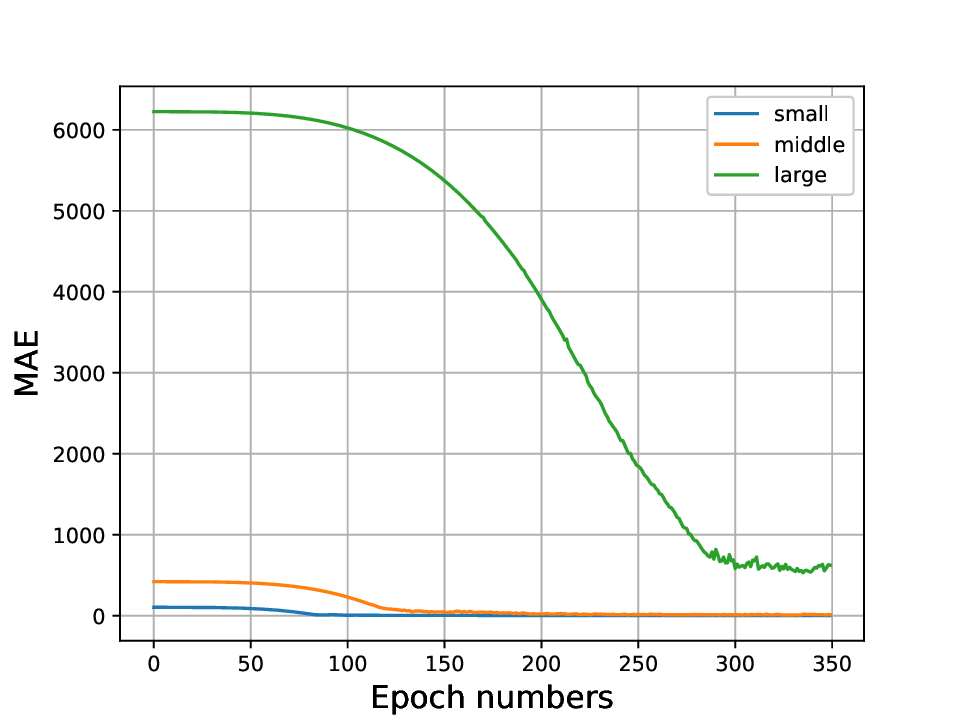}
        \put(68,54){aug-V}
      \end{overpic}
      \end{minipage}\label{Fig_LSTM-4}
      }
      \caption{The calculated MAE in the training process with "original" (upper) and "aug-V" (bottom) bi-LSTM models, respectively.}
      \label{Fig_LSTM-MAE}
      \end{figure}

\begin{figure}[htbp]
        \centering
        \subfigure[Different basis sets]{
        \begin{minipage}[t]{0.475\linewidth}
        \begin{overpic}[scale=0.50,bb=0 0 400 310]{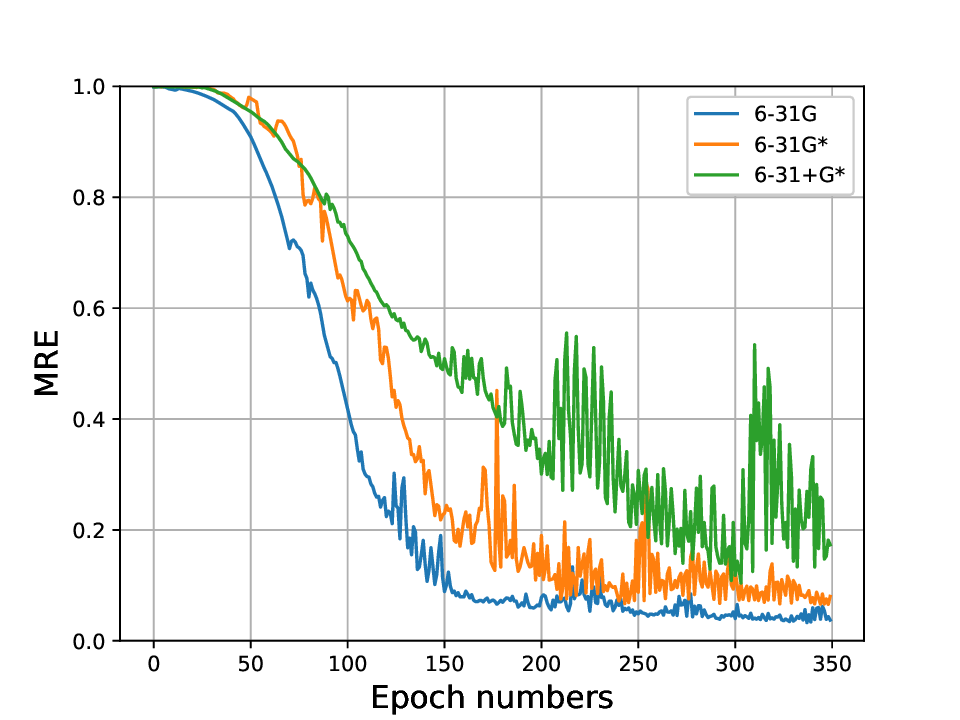}
          \put(50,64){original}
        \end{overpic}
        \begin{overpic}[scale=0.50,bb=0 0 400 310]{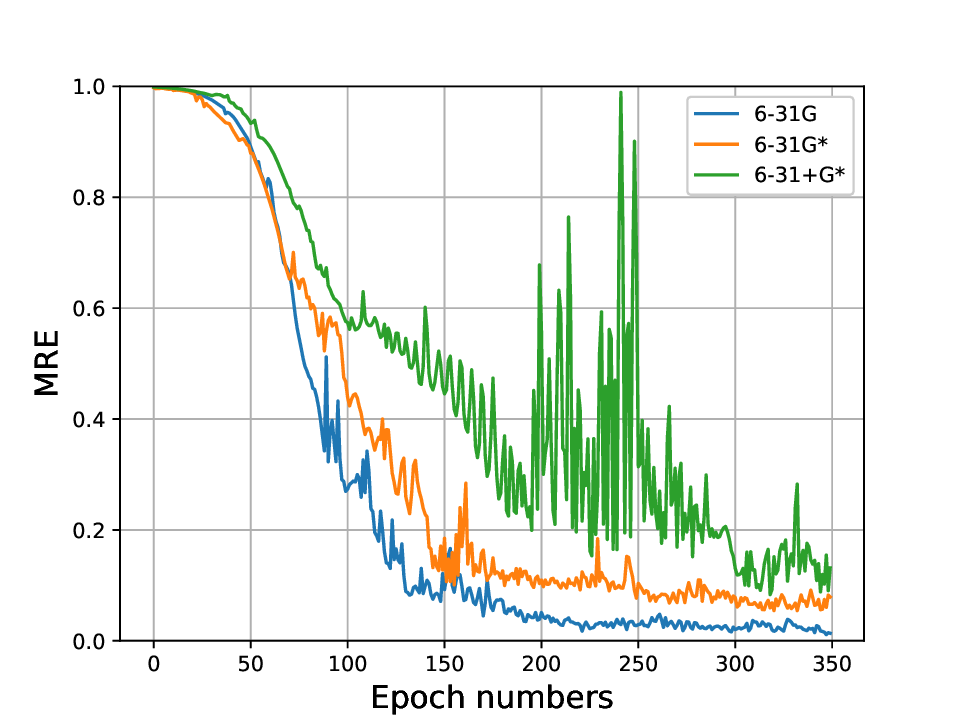}
          \put(50,64){aug-V}
        \end{overpic}
        \begin{overpic}[scale=0.50,bb=0 0 400 310]{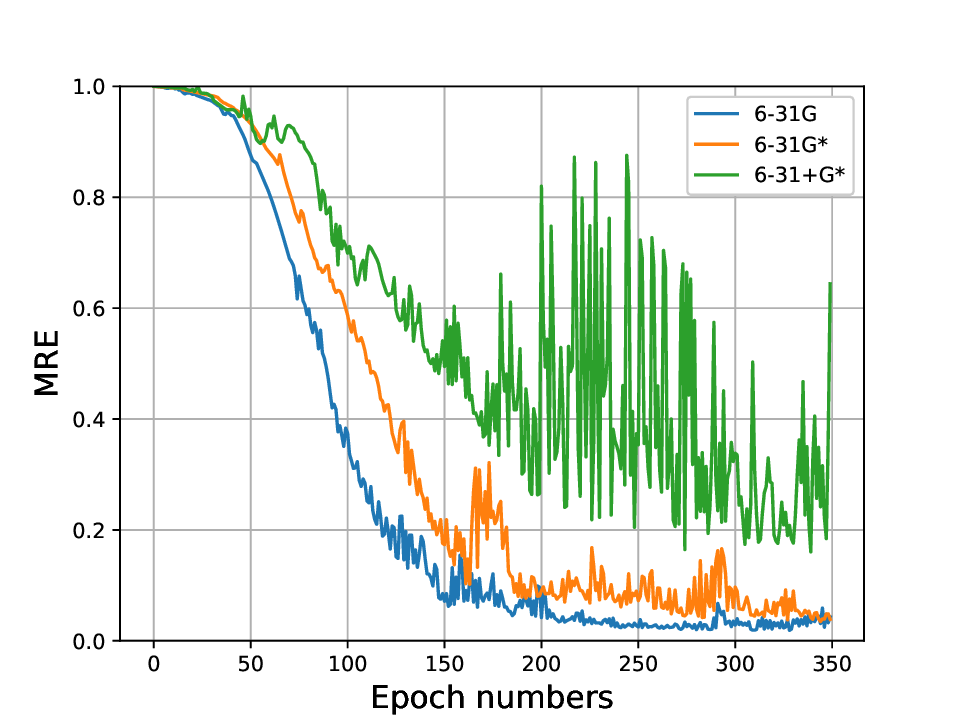}
          \put(50,69){aug-M}
        \end{overpic}
        \end{minipage}\label{Fig_MPNN-1} 
        }
        \subfigure[Different molecular weights]{
        \begin{minipage}[t]{0.475\linewidth}
        \begin{overpic}[scale=0.50,bb=0 0 400 310]{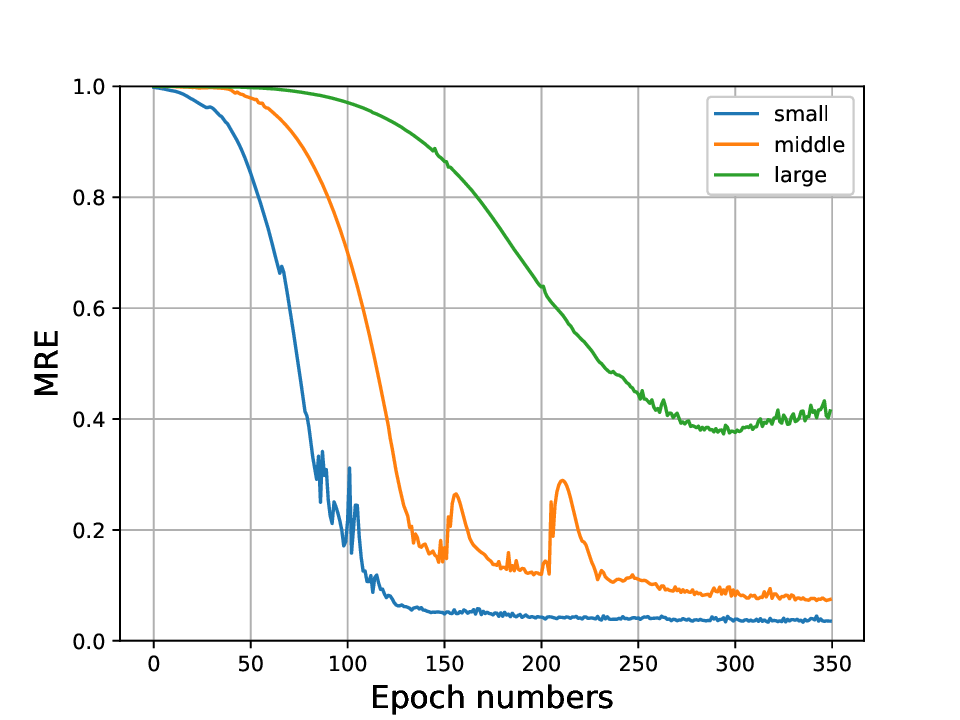}
          \put(59,64){original}
        \end{overpic}
        \begin{overpic}[scale=0.50,bb=0 0 400 310]{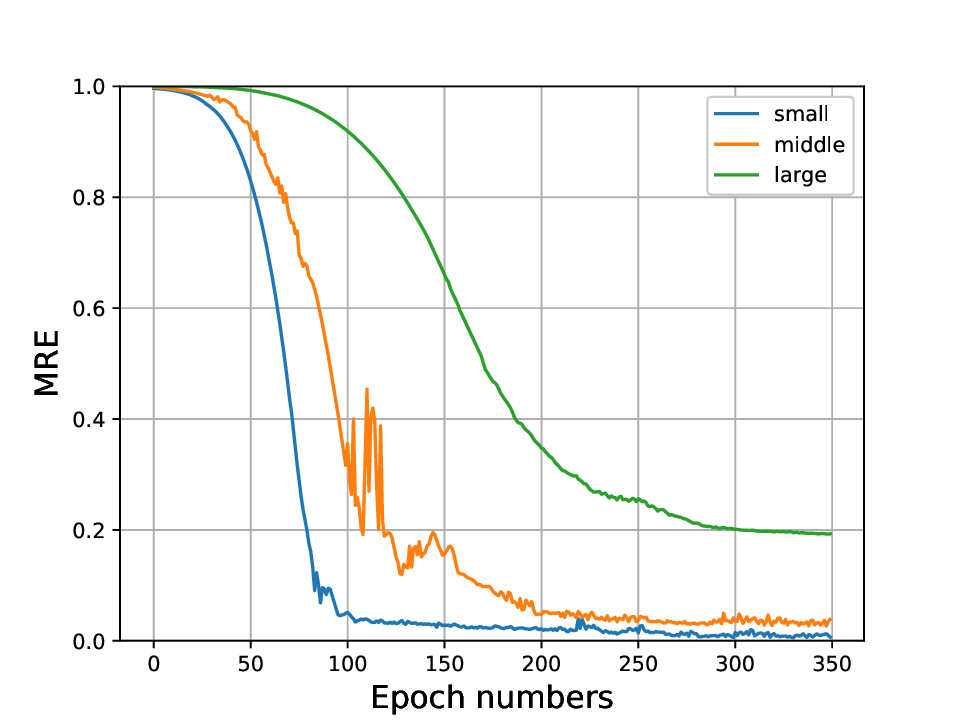}
          \put(59,64){aug-V}
        \end{overpic}
        \begin{overpic}[scale=0.50,bb=0 0 400 310]{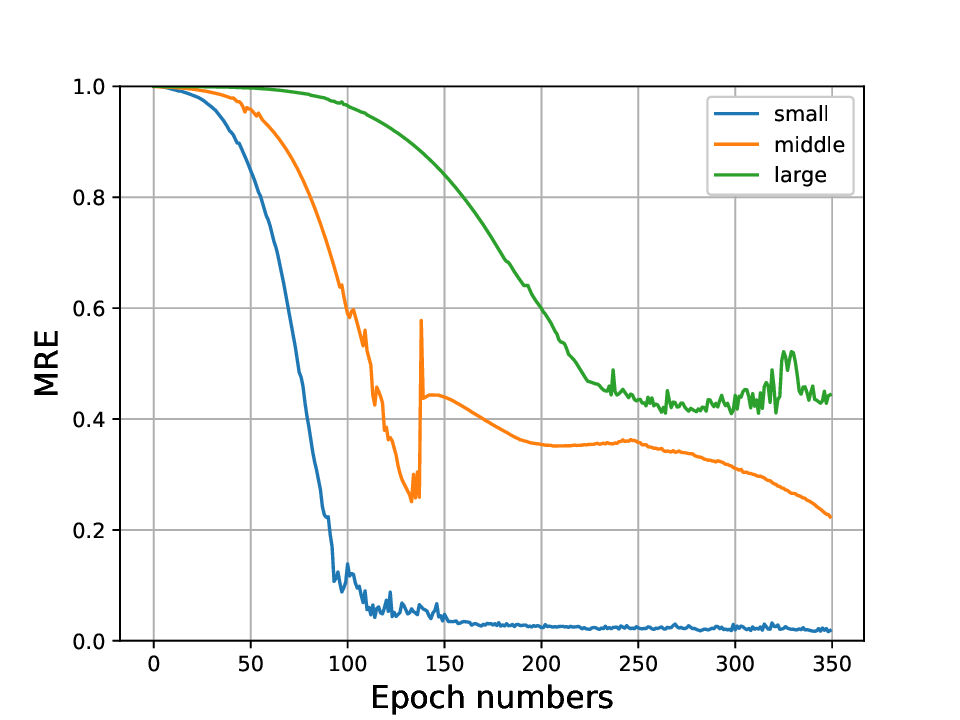}
          \put(59,64){aug-M}
        \end{overpic}
        \end{minipage}\label{Fig_MPNN-2}
        }
        \caption{The calculated MRE in the training process with "original" (upper), "aug-V" (middle), and "aug-M" (bottom) MPNN models, separately.}
        \label{Fig_MPNN-MRE}
        \end{figure}

\begin{figure}[htbp]
      \centering
      \subfigure[Different basis sets]{
      \begin{minipage}[t]{0.475\linewidth}
      \begin{overpic}[scale=0.50,bb=0 0 400 310]{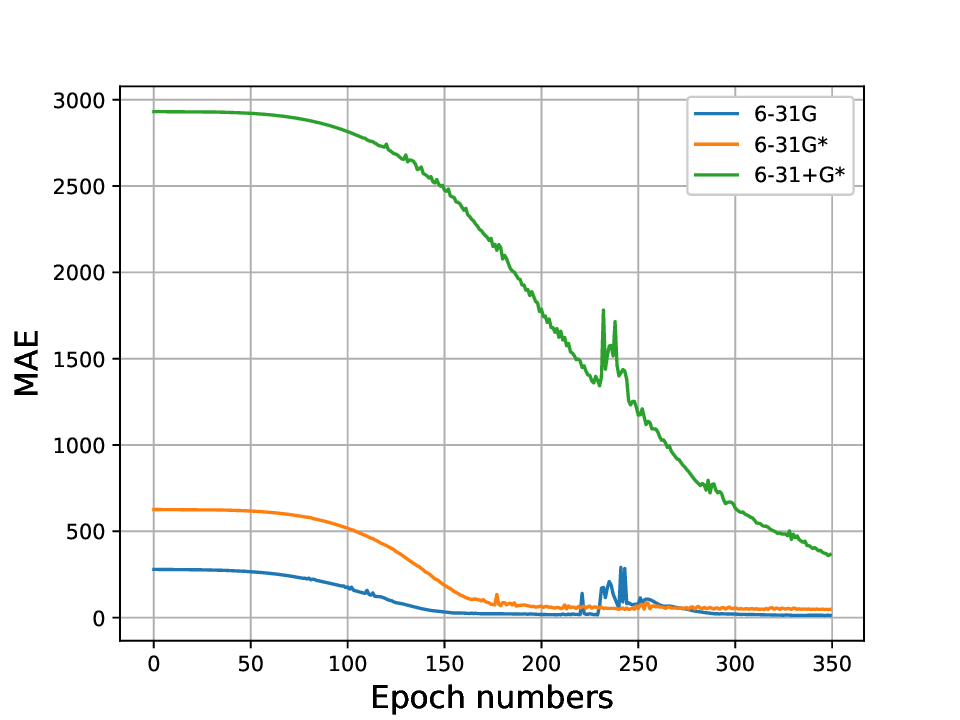}
        \put(50,64){original}
      \end{overpic}
      \begin{overpic}[scale=0.50,bb=0 0 400 310]{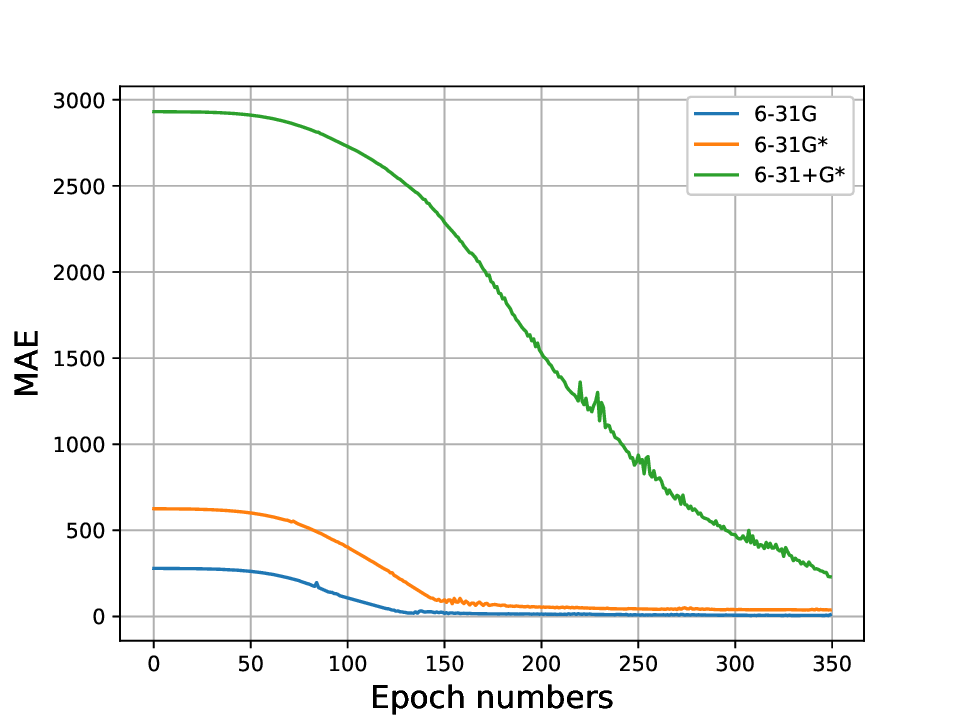}
        \put(50,64){aug-V}
      \end{overpic}
      \begin{overpic}[scale=0.50,bb=0 0 400 310]{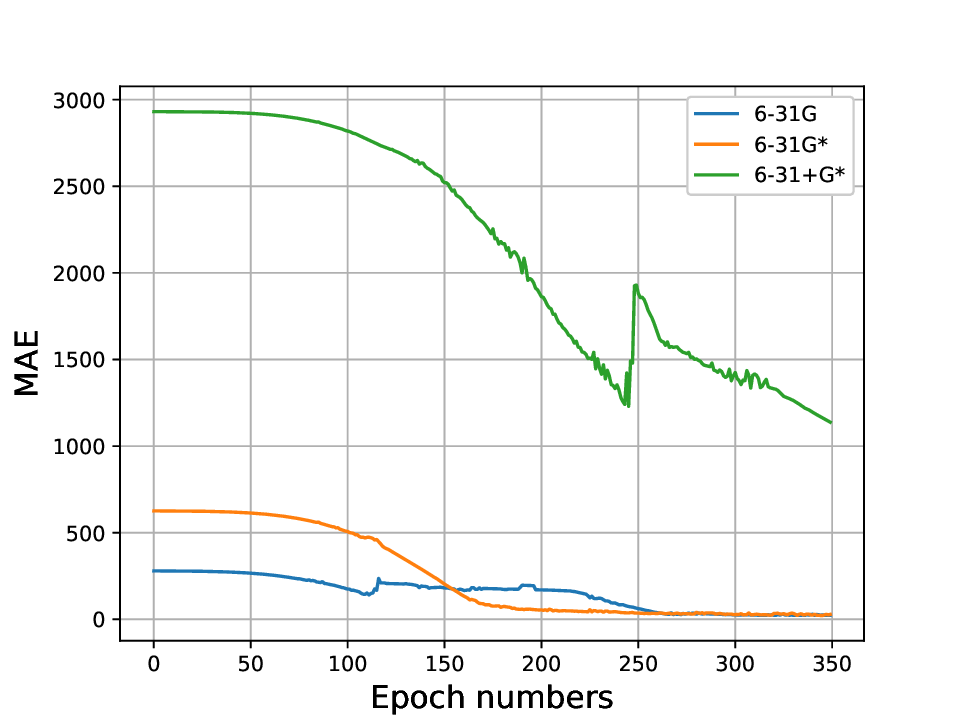}
        \put(50,69){aug-M}
      \end{overpic}
      \end{minipage}\label{Fig_MPNN-3} 
      }
      \subfigure[Different molecular weights]{
      \begin{minipage}[t]{0.475\linewidth}
      \begin{overpic}[scale=0.50,bb=0 0 400 310]{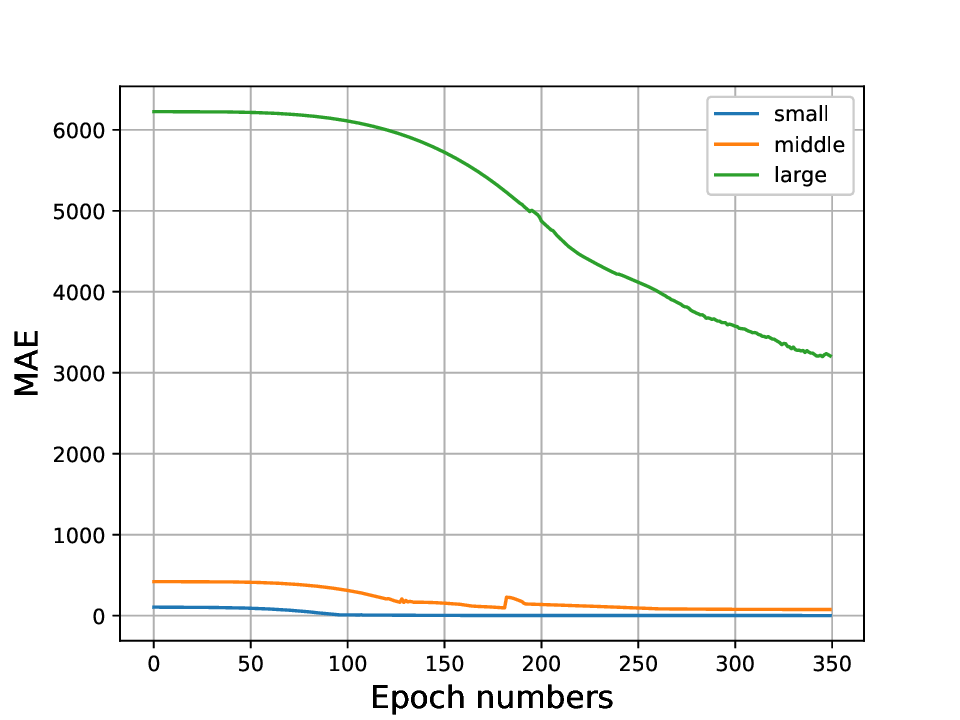}
        \put(59,64){original}
      \end{overpic}
      \begin{overpic}[scale=0.50,bb=0 0 400 310]{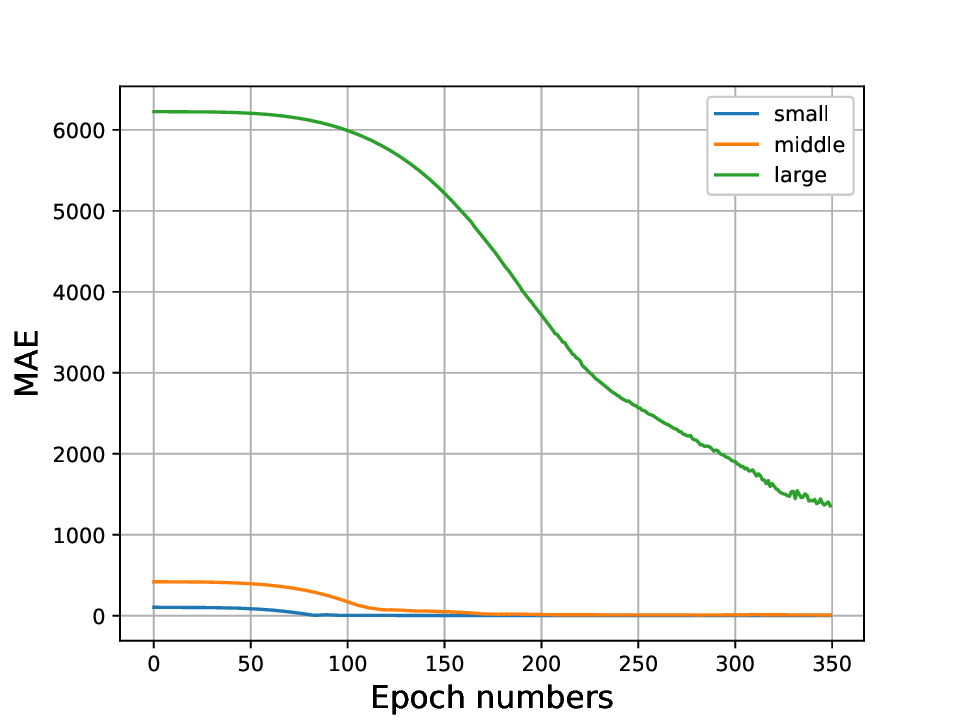}
        \put(59,64){aug-V}
      \end{overpic}
      \begin{overpic}[scale=0.50,bb=0 0 400 310]{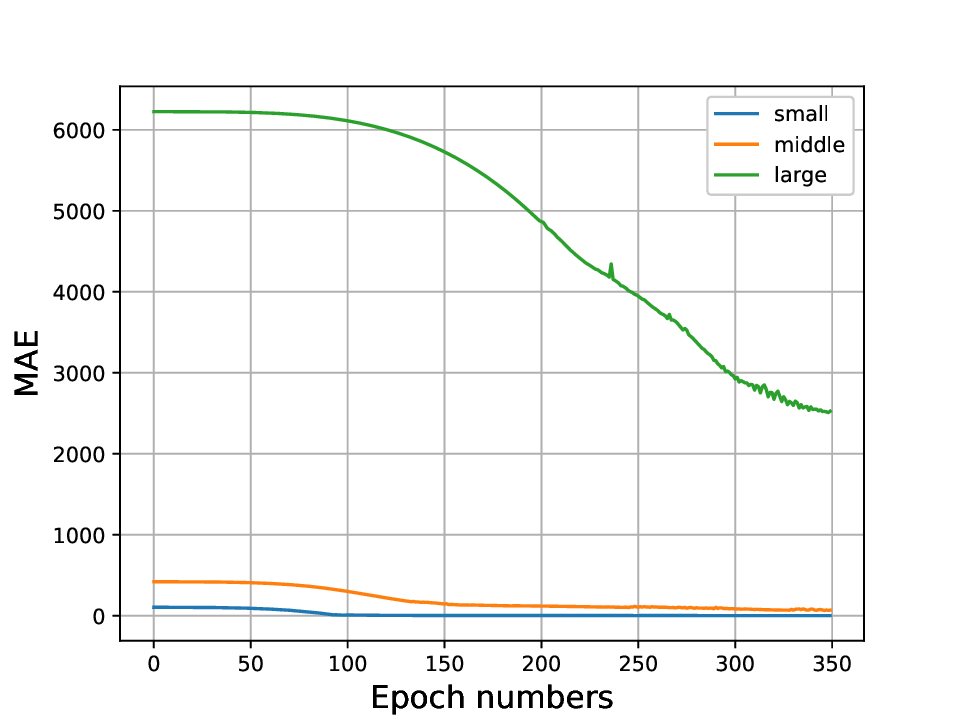}
        \put(59,64){aug-M}
      \end{overpic}
      \end{minipage}\label{Fig_MPNN-4}
      }
      \caption{The calculated MAE in the training process with "original" (upper), "aug-V" (middle), and "aug-M" (bottom) MPNN models, separately.}
      \label{Fig_MPNN-MAE}
      \end{figure}

\textcolor{blue}{MRE (Mean Relative Error) and MAE (Mean Absolute Error) are selected as evaluation metrics. MRE represents the average relative difference between the predicted and actual values, indicating the relative error. On the other hand, MAE is the average absolute difference, which is robust against outliers and offers a better reflection of errors in the actual scenario.
\begin{align*}
  MRE &= \frac{1}{n} \sum_{i=1}^{n} |\frac{y_i - \hat{y_i}}{y_i}| \\
  MAE &= \frac{1}{n} \sum_{i=1}^{n} |y_i - \hat{y_i}|
\end{align*}
}
The {MRE and MAE} results of the bi-LSTM model with the "aug-V" treatment are {presented} in Fig.\ref{Fig_LSTM-MRE} and Fig.\ref{Fig_LSTM-MAE}, respectively. 
In Fig.\ref{Fig_LSTM-MRE}, {it is observed that there is minimal variation compared to the original model, regardless of the basis sets or system sizes. 
This consistency is also reflected in Fig.\ref{Fig_LSTM-MAE}.}
The attention mechanism in the bi-LSTM model is applied to the molecular structure features, giving the features a higher weight, so that the effect of the added base group vector features is not obvious.
{In contrast to the bi-LSTM model, significant enhancements are evident in the graph-based models. }
The Fig.\ref{Fig_MPNN-1} shows the results of MPNN models with "aug-V and "aug-M" amendments. 
{The MPNN model with "aug-V" showcases superior performance compared to the original model, irrespective of the basis sets (Fig.\ref{Fig_MPNN-1}) or system sizes (Fig.\ref{Fig_MPNN-2})}. 
{The MREs exhibit enhanced convergence behavior during epoch iterations in this scenario.}
{On the other hand, the MPNN model with "aug-M" demonstrates minimal changes in its curves compared to the original model, as indicated by both the MRE results in Fig.\ref{Fig_MPNN-MRE} and MAE results in Fig.\ref{Fig_MPNN-MAE}. }
{The limited alteration may stem from the "aug-M" merging new features with existing molecular features and processing them across four levels, potentially leading to feature data losing its distinctiveness. 
This process underscores that the molecular feature data primarily aids in identifying molecular structures in calculations rather than focusing on the specifics of} AOs information within a given chemical space.

 
\subsubsection{\sffamily \large Improved ML-assisted static load balancing}

\begin{figure}[htbp]{
\centering
\includegraphics[scale=1.00,bb=00 0 400 230]{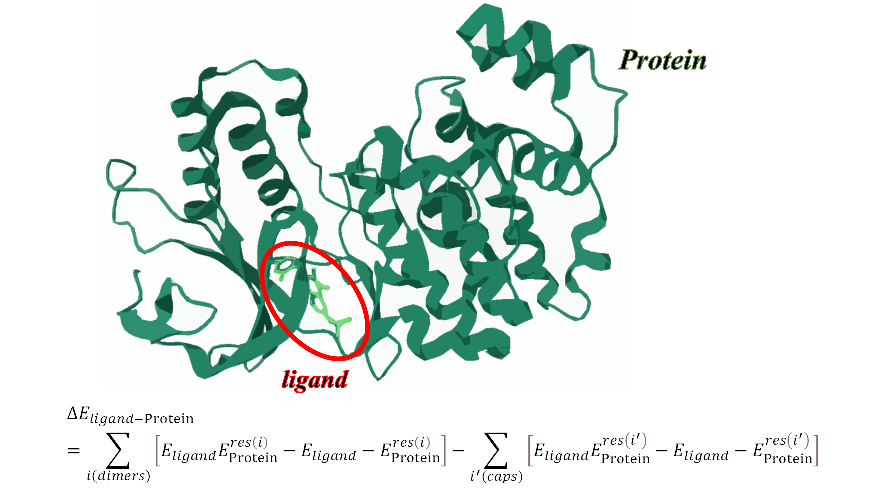}
\caption{The P38 protein (PDB ID: 3FLY) is used when benchmarking various SLB schemes, and in checking the fault tolerance of coded MFCC calculations.}\label{Fig_MFCC}
}
\end{figure}

\begin{figure}[htbp]{
\centering
\includegraphics[scale=0.55,bb=00 0 750 350]{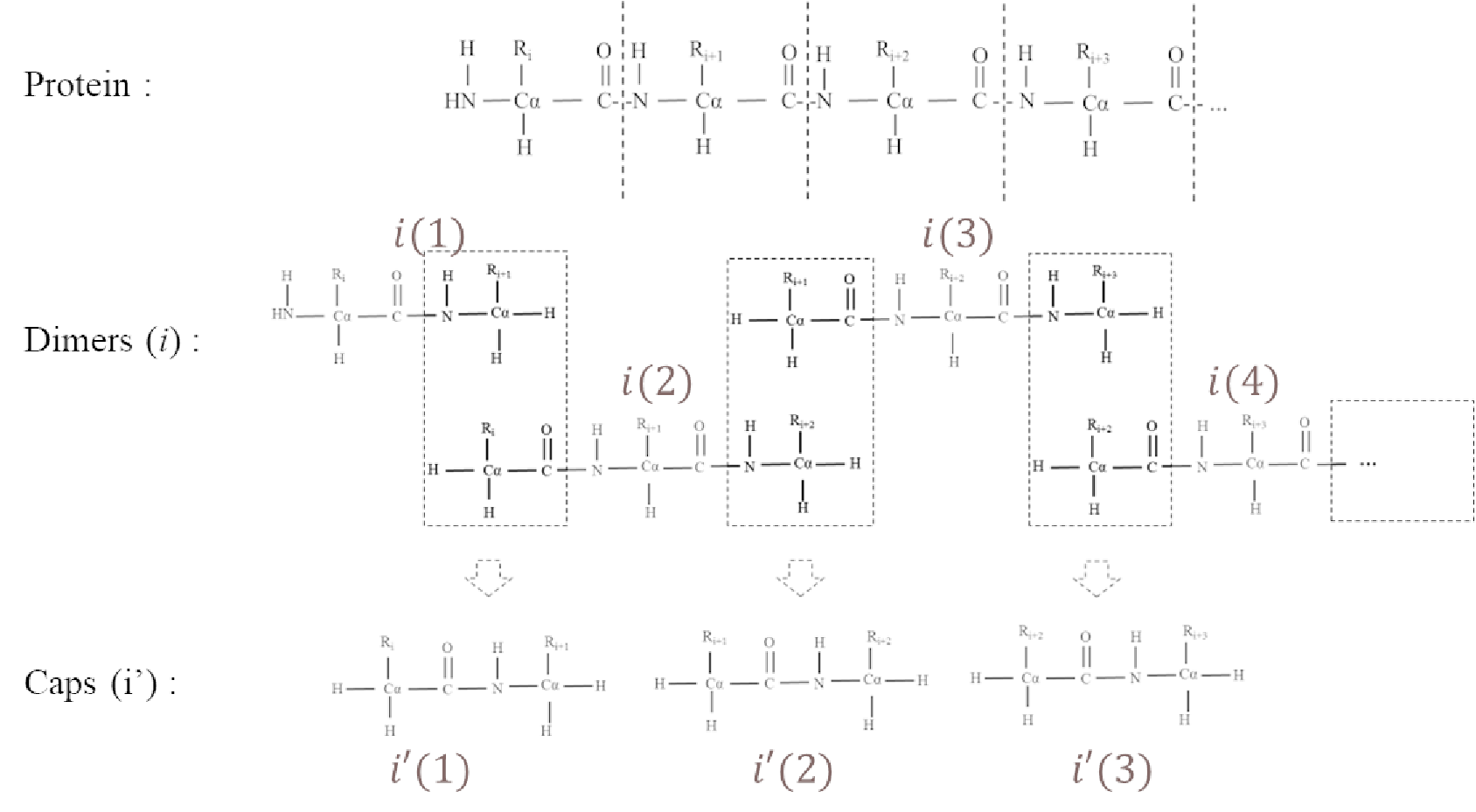}
\caption{{The illustration of MFCC fragmentation scheme for P38 protein.}}\label{Fig_MFCCindex}
}
\end{figure}

In Fig.9 of Reference \citenum{ma2023machine}, the "ML-assisted SLB + DLB" scheme demonstrates the highest computational efficiency, with SLB significantly enhancing the overall computational efficiency in distributed MFCC calculations of the P38 protein. 
{With the improved ML models, the same P38 system depicted in Fig.\ref{Fig_MFCC} and Fig.\ref{Fig_MFCCindex} can serve as the theoretical benchmark to evaluate the performance of the enhanced models in scheduling.
To isolate the contributions from SLB without the influence of DLB, only various SLB schemes are utilized in the benchmarking process.}

The default training suits in {\sc Fcst\_sys}\cite{ma2021forecasting} are {utilized for} training ML models, {with the "aug-V" variants serving as improved ML models.}
These SLB schemes are denoted as SLB(LSTM, original, {\sc{Fcst\_sys}}), SLB(LSTM, aug-V, {\sc{Fcst\_sys}}), and SLB(MPNN, aug-V, {\sc{Fcst\_sys}}), separately. 
Moreover, system-specific training sets, {particularly for protein systems}, are {employed alongside} the same "aug-V" variants, denoted as SLB(LSTM, original, {\sc{protein}}), SLB(LSTM, aug-V, {\sc{protein}}), and SLB(MPNN, aug-V, {\sc{protein}}) {correspondingly}.

\begin{figure}[htbp]
  \centering
  \begin{overpic}[scale=0.3750,bb=10 0 1400 700]{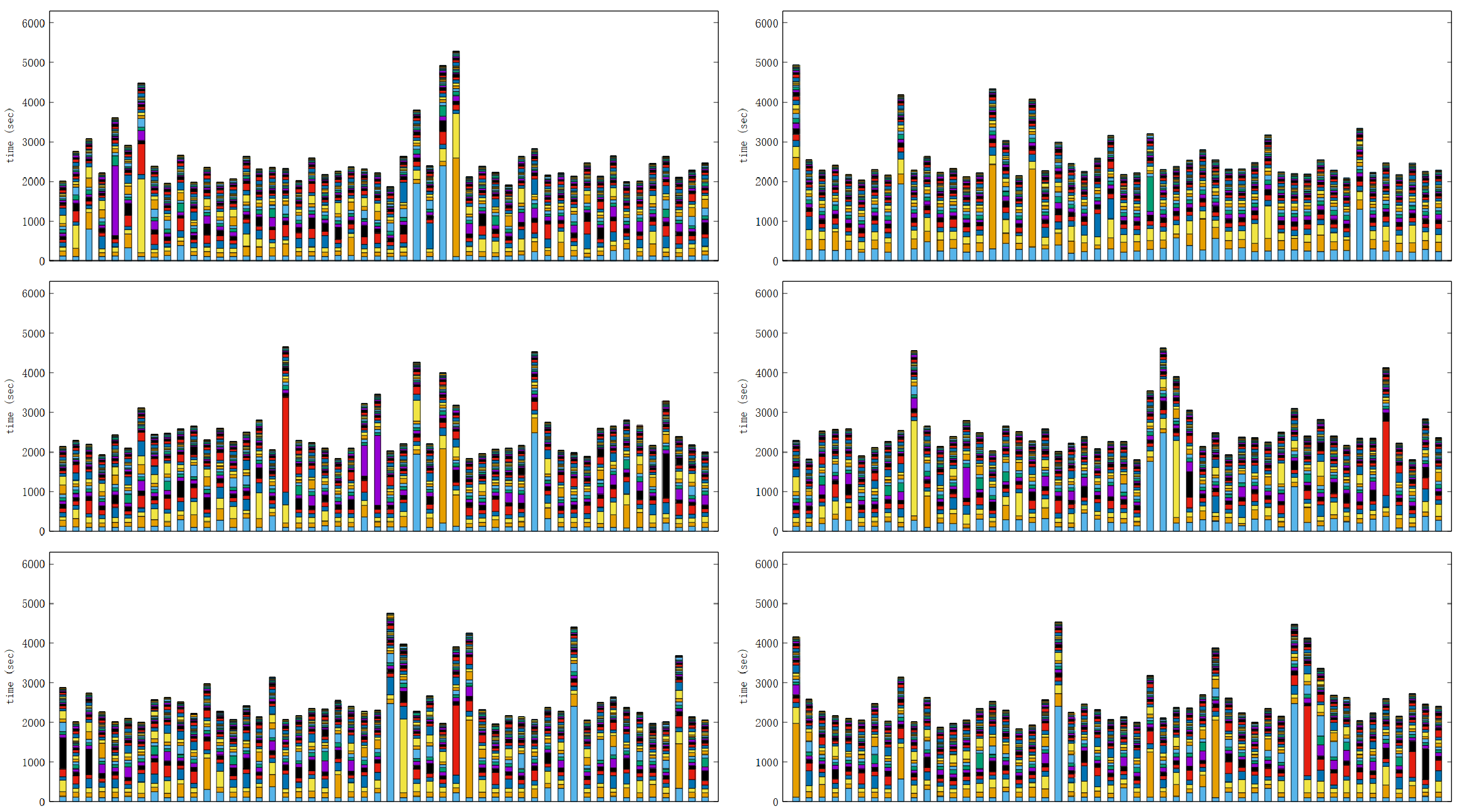}
    \put(40,47.5){(A)}
    \put(40,30.5){(B)}
    \put(40,13.5){(C)}
    \put(50,47.5){(D)}
    \put(50,30.5){(E)}
    \put(50,13.5){(F)}
\end{overpic}
  \caption{The illustration of elapsed times for M06-2x/6-31G** snapshot of P38 protein using MFCC computational approach in the distributed case with 50 HPC nodes. {In each picture, different colors present different task; and each bar is independent, whose length shows the time of calculating. The x axis means situation on different nodes.} Different scheduling schemes are employed, they are (A) SLB(LSTM, original, {\sc{Fcst\_sys}}), (B) SLB(LSTM, aug-V, {\sc{Fcst\_sys}}), (C) SLB(MPNN, aug-V, {\sc{Fcst\_sys}}), (D) SLB(LSTM, original, {\sc{protein}}), (E) SLB(LSTM, aug-V, {\sc{protein}}), (F) SLB(MPNN, aug-V, {\sc{protein}}), separately. The (A), (D) are from Ref.\citenum{ma2023machine}.}
  \vfill
  \label{Fig_MLSLB}
  \end{figure}

\begin{figure}[htbp]
  \centering
  \includegraphics[scale=0.90, bb = 0 0 375 250]{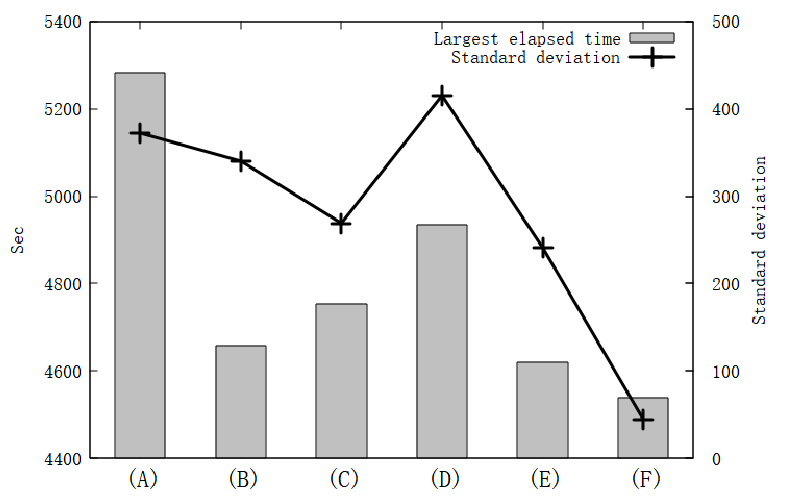}
  \caption{The largest elapsed time and the standard deviation for all the distributed HPC nodes when using different SLB schemes.}
  \vfill
  \label{Fig_time}
  \end{figure}

All the results of elapsed times using different SLB schemes with 50 distributed nodes are depicted in Fig.\ref{Fig_MLSLB}, {while the analysis based on the largest elapsed time and standard deviation is presented in Fig.\ref{Fig_time}.}
{The analysis shows} that both the system-specific ML models and the "aug-V" variants {contribute to improved} task distribution. 
The best combination is SLB(MPNN, aug-V, {\sc{protein}}) scheme, in which the largest elapsed time in node is 4539 sec. with a small standard deviation.

\subsubsection{\sffamily \large Coded quantum chemical calculations}

The coded quantum chemical calculation {employs} the MFCC and REM framework. 
The calculation of ligand-protein interaction of P38 protein (PDB ID: 3FLY), which is illustrated in Fig.\ref{Fig_MFCC}, is used as the example in checking the fault tolerance of coded calculation. It can be noticed that the binding/interaction energy can be treated as the ligand-dimer interaction minus the ligand-cap interaction, so that the no-redundant fragments in most calculations are the $E_{ligand} E^{res(i)}_{Protein}$, $E^{res(i)}_{Protein}$, $E_{ligand} E^{res(i')}_{Protein}$, and $E^{res(i')}_{Protein}$, separately. Notice that the $i$ and $i'$ indexes the dimers and caps, respectively, and the no-redundant 4 fragments with same index of i/i' can be treated as a minimum group unit for calculating protein-ligand interactions. 

Basing on the idea of gradient coding, the task number $n$ can be set to 4, assuming the tolerable straggler node $s$ is 1, then the $B$ matrix listed in subsection 2.2.2 can be used as the encoding matrix. 
The decoded binding energies {are} compared with the reference energy obtained in Ref.\citenum{ma2023machine}, and a few randomly selected results are presented in Table \ref{table_1}.
For the fragment group with i/i' as 0029, the encoding computations yield 4 identical results, indicating that all 4 computing nodes have correctly obtained the results. 
{Conversely,} for the fragment groups with i/i' as 0062, calculations for one set of data on one node exceeded the threshold we set, {leading to a lack of returned result.} 
However, {the correct result can still be decoded from the remaining three computing nodes.}{\cite{li2024coded}}

\begin{table}
  \centering
  \caption{Checking the fault tolerance in the coded MFCC calculations for the binding energies in the randomly selected fragment groups.}
  \begin{tabular}{c c c} 
  \toprule  
  Fragments group (i/i')  & Reference & Decoded binding energy(a.u)\\
  \midrule  
  0029& $9.29399\times 10^{-5}$
      &\begin{tabular}{c} 
          $9.29399\times 10^{-5}$ \\ 
          \cline{1-1}
          $9.29399\times 10^{-5}$ \\
          \cline{1-1}
          $9.29399\times 10^{-5}$ \\
          \cline{1-1}
          $9.29399\times 10^{-5}$ \\
      \end{tabular}\\
          \hline
  0062& $-6.86651\times 10^{-5}$ 
      &\begin{tabular}{c} 
          $-6.86651\times 10^{-5}$ \\ 
          \cline{1-1}
          (no result) \\ 
          \cline{1-1}
          (no result) \\ 
          \cline{1-1}
          (no result) \\ 
      \end{tabular}\\
  
  \bottomrule 
  \end{tabular}
  \label{table_1}
\end{table}

\begin{figure}[htbp]
  \centering 
   \includegraphics[scale=1.75,bb=0 0 100 65]{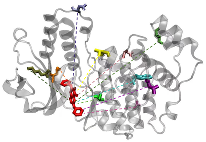}
   \includegraphics[scale=2.75,bb=0 0 100 65]{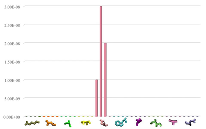}
  \caption{Derivations that may introduced by the encoding-decoding process are evaluated. Several fragment groups in coded calculating protein-ligand interactions are illustrated, and the derivations between results from encoding-decoding process and these from Ref.\citenum{ma2023machine} are also shown.} \label{Fig_codedMFCCcheck}
\end{figure}

{After completing the fault tolerance assessments}, the errors that may introduced by the encoding-decoding process are also checked.  
This time, all sampled fragment groups, {along with the discrepancies between the results obtained} from encoding-decoding and those from {Ref.\citenum{ma2023machine}}, are {depicted} in Fig. \ref{Fig_codedMFCCcheck}.
{It is evident that the majority of discrepancies resulting from encoding-decoding are less than $1.0\times10^{-10}$, with the largest discrepancy being at the $1.0\times10^{-8}$ level.} 
Actually, the $3.0\times10^{-8}$, $2.0\times10^{-8}$, $1.0\times10^{-8}$, and $1.0\times10^{-10}$ derivations for this specific coded fragments' group may also reflect the very tiny convergence mismatch. {\cite{li2024coded}}

\subsection{\sffamily \Large Applying in fragmented-based excited-state calculations}

{After completing} the term-wise assessments for all components (improved ML models, scheduling optimization, coded computing, etc.) that may {aid} practical quantum chemical calculations, we {proceed} to the calculation of excited states with one or several excitable centers. 
The green fluorescent protein anion (GFPA) dye in water solvent and drug-water mixed molecular clusters are used as benchmark systems, respectively.

\subsubsection{\sffamily \large Absorption spectra of GFPA in water solvent}
Before {calculating} the excited states of the GFP solvent model, reasonable structures should first be obtained using molecular dynamics (MD) simulations .\cite{zuehlsdorff2018combining} 
Herein, the generalized AMBER force field (GAFF) \cite{wang2004development} is applied to the solute, while water is described using the TIP3P model .\cite{jorgensen1983comparison} 
For the central dyes, the {\sc ANTECHAMBER} program \cite{wang2004development} is used to generate appropriate force field parameters, with input structures optimized in vacuum at the M06-2x/6-31G* level. 
The MD simulations are performed using the {\sc Amber} software, employing the Langevin dynamics approach with a collision frequency of 1 $ps^{-1}$. The simulation begins at 0K, then targets a temperature of 300 K, taking several snapshots at this temperature.

Once the snapshot is obtained, the 100 water molecules {closest} to the central GFPA dye are selected, and all monomers within an inter-distance of 4\AA \ are chosen as dimers. 
If any dimers contain the central GFPA dye molecule, then both the dimers and the related monomers (water molecules and GFPA dye) are encoded during the excited states calculations using the TDDFT approach. Both the decoded and encoded subsystems are handled together by the {\sc ParaEngine} package for convenience. \cite{ma2023machine} After {completing} the calculations, the REM-TDDFT approach {is used to obtain} the excited states for the whole system. The results for a sampled snapshot are shown in Table \ref{LBresults2}. It is {observed} that the absorption is mainly located in the GFPA molecule {within} the solvent system, which {aligns with} the results of previous studies.\cite{zuehlsdorff2018combining}

Nevertheless, {it should be noted} that the encoding-decoding process can be applied to any observable parameters. 
In the previous MFCC calculation of binding energy, the coded fragments {had} specific physical meanings, i.e., the binding energy from fragment group units. Actually, any related observables, e.g. excitation energies here, can be {used} as markers in the coded {process}. 
For example, dimers containing the GFPA molecule and specific water molecules within these dimers can be coded as a fragment group, upon which the $B$ matrix can be generated.

\begin{figure}[htbp]{
\centering
\includegraphics[scale=1.00,bb=0 0 250 250]{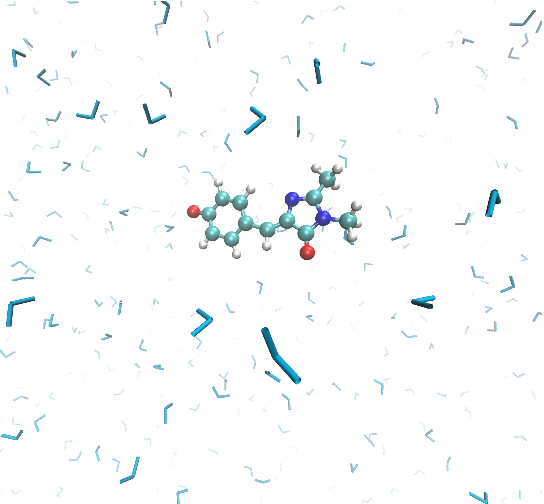}
\caption{The illustration of GFPA in water solvent.}\label{Fig_GFPA}
}
\end{figure}

\subsubsection{\sffamily \large Low-lying excited states for clusters with several excitable centers}

\begin{figure}[htbp]{
\centering
\begin{overpic}[scale=0.75,bb=00 0 200 200]{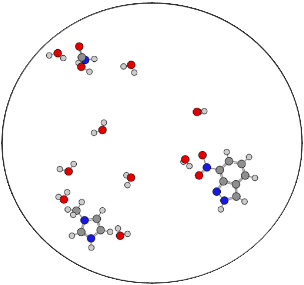}
  \put(24,75){Cluster-1}
\end{overpic}
\begin{overpic}[scale=1.00,bb=00 0 200 200]{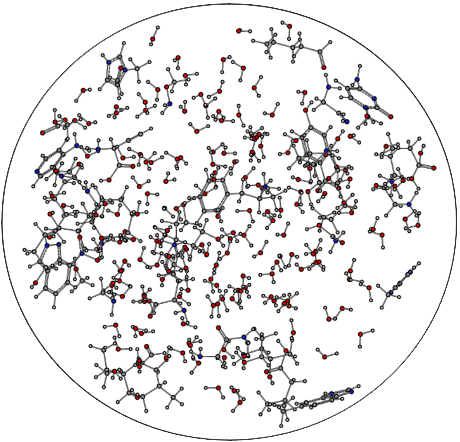}
  \put(-20,74){Cluster-2}
\end{overpic}
\caption{The illustration of two molecular clusters with several excitable centers.}\label{Fig_exrem}
}
\end{figure}

{In the final calculations}, clusters with several excitable centers are used. 
The cluster models were {created using the} {\sc PackMol} package \cite{martinez2009packmol}, {incorporating} water and drug molecules from the {DrugBank} library .\cite{wishart2018drugbank} 
This system was also {utilized} in our previous work {to illustrate} 'ML-assisted SLB + DLB' solvation effect calculations. {In this study, we also consider low-lying excited states that extend beyond ground state interactions.}
{The system consists of 203 monomers}, including various drug molecules and water. 
{Since} the cluster is too large to {obtain} a reliable reference, a reduced cluster is {selected} by extracting some monomers from the original cluster. 
{Both molecular clusters} (denoted as Cluster-1 and Cluster-2) are illustrated in Fig. \ref{Fig_exrem}.

For the reduced molecular cluster, both the coded REM-TDDFT and the normal TDDFT calculations can be {performed using} {\sc ParaEngine} and {\sc Gaussian09}, \cite{g09} respectively. 
The calculated results are listed in Table \ref{LBresults2}. It can be {observed} that the excitation energies for these two clusters are {nearly identical, primarily} stemming from a specific monomer. 
However, the excitable monomers can be identified for more than this one specific monomer for both REM-TDDFT and TDDFT approaches in Cluster-1 systems. 
The wave functions from REM-TDDFT imply that more excitable monomers are involved in the more complicated Cluster-2 system.

\begin{table}[htbp]	
\centering
	\caption{Calculated excitation energies of the lowest singlet excited states of the two cluster systems with REM-TDDFT and TDDFT approaches. The possible excited monomers are also listed for REM-TDDFT calculations with their coefficients are shown in brackets.} 
	\label{LBresults2}		
	\begin{tabular}{ccccccccccccccccccccccc}
		\hline
		\hline
\multirow{1}*{}  & & \multicolumn{2}{c}{{\sc REM-TDDFT}} && &   \multicolumn{1}{c}{ {\sc TDDFT} }   \\
                \cline{3-5}   \cline{7-8} 
                 & &   eV  &    excitable monomers' ID (weight$^a$)    &&  &   eV     \\        
		\hline   
	  \vspace{0.1cm}	
  \textbf{GFPA}    &&    &    &&  &      &     &   \\
    \vspace{0.1cm}
   \textit{ $S_1$}  && 3.104  &  1(1)   &&  &   {3.096}      \\  
  \textbf{Cluster-1}    &&    &    &&  &      &     &   \\
    \vspace{0.1cm}
   \textit{ $S_1$}  && 2.518  &  12(0.76) \  \ 9(0.65)   &&  &  2.509         \\  
     \vspace{0.1cm}
  \textbf{Cluster-2}                &&       &      &&  &     &         \\ 
    \vspace{0.1cm}
    \textit{ $S_1$ }   &&  2.517 & 141(0.21) \  138(0.39) \ 113(0.47)  &&  &  \multirow{1}*{-$^b$}   \\  
                       &&        & 110(0.39)  \ 65(0.39)  \ 46(0.33)  \ 6(0.39)  \\                  
       \hline
       \hline
	\end{tabular}
    \begin{tablenotes}
      \scriptsize
      \item \qquad {a : $C_I$ value in Eq.(\ref{eqProj})}
      \item \qquad {b : not available}
    \end{tablenotes}
\end{table}

\section{\sffamily \Large CONCLUSIONS}

Basing on recent work in ML-assisted scheduling optimization, we further propose
1) improved ML models for better predictions of computational loads;
2) {the concept of} coded computation to introduce fault tolerance during distributed calculations;
3) {applications of these improvements within the REM framework} for calculating excited states.
In the improved ML models, the descriptor for the basis sets is extended from a point to a vector (aug-V) or matrix (aug-M), {resulting in significant accuracy improvements}. For {instance, there is} about  50\% reduction in MRE values when comparing the predicted results of MPNN(aug-V) to these of MPNN(original). 
{In coded computing, gradient coding is introduced to provide fault tolerance during distributed calculations, allowing the detection or correction of abnormal results caused by convergence issues or anomalous nodes.}
Both the improved ML models and the coded computing technique can be integrated in the {\sc ParaEngine} package, {where the REM-TDDFT utility is already} implemented for describing the excited states of large molecules and molecular clusters.

{The benchmark calculations include the P38 protein and a solvent model with one or several excitable centers.}
The results {demonstrate} that both computational efficiency and capacity of fault tolerance can be improved. 
For instance, the SLB assignments can be at least improved by 10\% $\sim$ 15\% when the improved ML models are employed, and the coded computing can guarantee the abnormal results can be easily located with some of them can be corrected automatically. 
Additionally, their preliminary integration with MFCC and REM-TDDFT approaches show its potential usage in the ML-assisted automated fault-tolerant calculation for large systems.

Reaching the end, we should mention that the current implements are better work within HPC clusters. The cross-domain calculation is not implemented yet. Working on this direction is currently in progress in our laboratory. We hope the computational scheme can be finally evolved like {\sc Folding@Home}\cite{voelz2023folding} or {\sc Seti@Home},\cite{anderson2002seti} so that more computational resources can be utilized to accelerate the related research. 

\subsection*{\sffamily \large ACKNOWLEDGMENTS}
Y. Ma thank Prof. Haili Xiao for the en-lighting discussions of fault-tolerant. This work was supported by National Natural Science Foundation of China (Nos. 22173114, 22333003 to Y. Ma, and 42371476 to D. Guo), Strategic Priority Research Program (XDB0500101), Youth Innovation Promotion Association (No. YIPA2022168), Network and Information Foundation (CAS-WX2021SF-0103-02), Project of Computer Network Information Center (CNIC20230201) of Chinese Academy of Sciences, and the Fundamental Research Funds for the Central Universities (BUCTRC202132). K. Yuan also thank the support from College Students Innovative Practice Training Program of Chinese Academy of Sciences. Most of the computational experiments were implemented in the ”ORISE” and ”ERA” supercomputers, we are also highly appreciated the helps from the supporting team.

\subsection*{\sffamily \large Data Availability Statement}

The data that support the findings of this study are available from the corresponding author upon reasonable request.

\section{\sffamily \Large Appendix}

\subsection{\sffamily \large Example of ML-assisted Coded Scheduling Optimization}

{Herein, we take an example to explain how the scheduling is accomplished based on the $B$ matrix. 
Assuming a whole system calculation can be divided into 100 sub-system calculations that denoted as task 1-100, and every four tasks can be treated as a task group, e.g. 1-4, 5-8, as illustrated in prevision section and ref.\citenum{li2024coded}.  }

{If fault tolerance is {unnecessary, a greedy algorithm suffices to maintain load balancing among computing nodes.} \cite{ma2023machine} 
If fault tolerance is {required for certain subsystem calculations, such as tasks 1-4, 5-8, and 9-12, replicas will be created for each task from 1 to 12, and then these tasks will be encoded.} 
In fragmentation calculations, tasks 1 to 12 represent the 12 longest tasks among all, {as they are the interacting fragments (i.e., dimers). 
This approach also allows us to observe the time cost incurred when using an encoded distributed computing system in the worst-case scenario.}
To avoid computational overhead caused by excessively large encoding and decoding matrices, we have divided these 12 tasks into three groups: 1-4, 5-8, and 9-12. 
These three groups will {utilize} the same 4 $\times$ 4 encoding matrix. 
According to Algorithm \textbf{1}, {assuming $s = 1$ and $n = 4$, one of the valid $B$ matrices can be generated as follows},}

\begin{equation*}
B = 
\begin{bmatrix}
 1 & -1.42677270 & 0& 0 \\
 0 & 1 &  -7.65737406& 0 \\
 0 & 0 & 1&0.05106546 \\ 
 1.79241288 & 0 & 0 & 1\\
\end{bmatrix}
\end{equation*}

\begin{figure}[htbp]
   \centering
   \begin{overpic}[scale=0.95,bb=0 0 500 450]{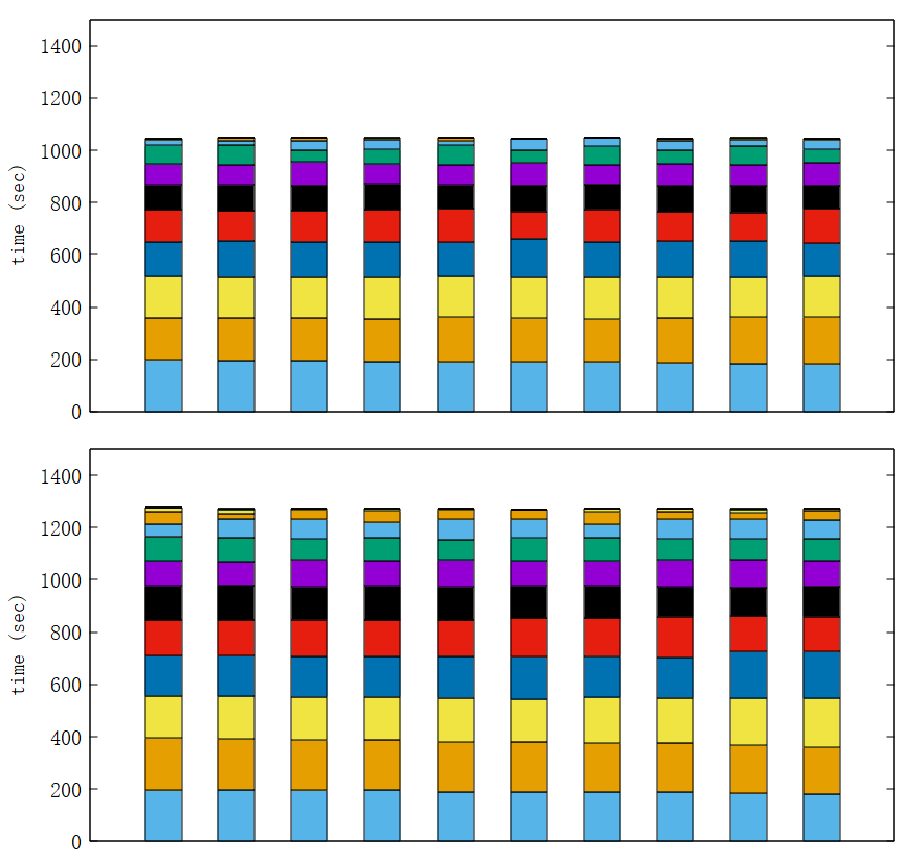}
   \end{overpic}
  \caption{ {The illustration of elapsed times for toy model in Appendix with partly coded (bottom) or without coded (upper) scheduling optimization using 10 HPC nodes. In each picture, different colors present different task; and each bar is independent, whose length shows the time of calculating. The x axis means situation on different nodes.}}
 \vfill
 \label{Fig_TimeContrast}
\end{figure}

\begin{figure}[htbp]
  \centering
  \begin{overpic}[scale=0.90,bb=0 0 450 350]{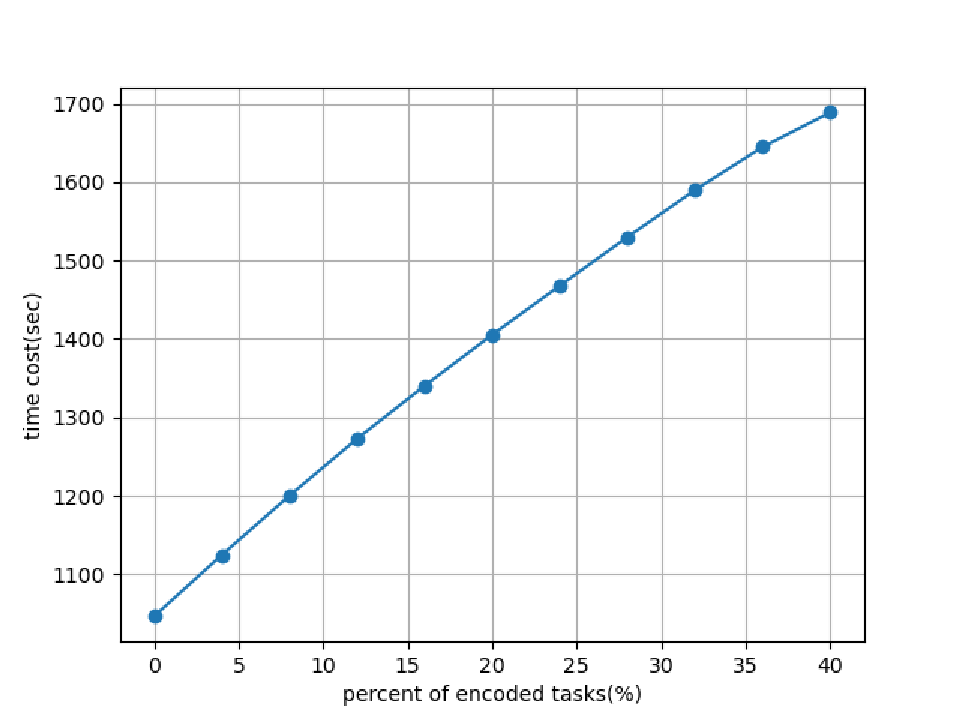}
  \end{overpic}
 \caption{ {The time consumption under different proportions of tasks to be encoded.}}
\vfill
\label{Fig_TimeCost}
\end{figure}

{In this $B$ matrix, the indices of non-zero elements in the $i$-th row indicate the task numbers that need to be computed on computing node-$i$. 
For instance, in the {second} row of the matrix above, tasks-{$2$} and task-{$3$} can be assigned to node-{$2$}, and the coefficients 1 and -7.65737406 are used to encode the results of these tasks. 
Once the encoding tasks are specified, SLB scheduling optimization for all tasks can be carried out using planning algorithms, such as the greedy algorithm in our previous work .\cite{ma2023machine} The optimized task scheduling is listed in Table \ref{table_3}, and the theoretical computational costs are illustrated in Fig.\ref{Fig_TimeContrast}. It can be observed that the partly coded approach only incurs about 20\% additional computational costs. 
{To further explore the correlation between the proportion of tasks to be encoded and the time consumed, we conducted calculations for various task proportions, as depicted in Fig.\ref{Fig_TimeCost}}. 
It should be feasible to search for the equilibrium point between efficiency and fault tolerance.}

\begin{table}
  \centering
  \caption{{Optimized task scheduling with and without coded fault-tolerance}}
  \begin{tabular}{c c c} 
  \toprule  
  Node& Task-list(no coded) & Task-list(coded)\\
  \midrule  
  node-1& 1 20 26 38 43 58 66 75 88 95      & 1.1 2.1 20 28 36 40 54 62 78 79 92   \\
  \hline
  node-2& 2 19 24 34 47 55 67 71 89 92    &2.2 3.1 19 27 35 39 58 59 75 88 89 100   \\
  \hline
  node-3& 3 18 25 35 46 54 62 80 83 93    &3.2 4.1 17 25 32 43 50 66 71 85 96    \\
  \hline
  node-4& 4 17 22 36 44 56 70 76 84 97 100      &1.2 4.2 18 26 34 41 55 61 76 81 94   \\
  \hline
  node-5& 5 14 28 39 42 59 68 72 90 91    &5.1 6.1 15 22 33 42 51 67 69 83 97    \\
  \hline
  node-6& 6 15 23 31 50 53 63 78 81    &6.2 7.1 16 21 30 44 56 63 74 84 98    \\
  \hline
  node-7& 7 16 21 37 45 57 69 73 86    &7.2 8.1 13 24 31 45 57 60 77 80 91  \\
  \hline
  node-8& 8 13 27 33 48 52 64 77 82 94      &5.2 8.2 14 23 29 46 52 65 70 86 93   \\
  \hline
  node-9& 9 12 30 32 49 51 65 74 87 98 99     &9.1 10.1 11.2 12.1 37 48 49 68 72 87 90 99  \\
  \hline
  node-10&10 11 29 40 41 60 61 79 85 96     &10.2 11.1 9.2 12.2 38 47 53 64 73 82 95    \\
  \bottomrule 
  \end{tabular}
  \label{table_3}
   \end{table}

\clearpage


\bibliography{references}

\clearpage

\end{document}